\documentclass[11pt,preprint]{aastex}
\slugcomment{Accepted for publication in {\it Astrophysical Journal}}
\shorttitle{Substellar Mass Function of Open Clusters}
\begin{document}

\title{Substellar Mass Function of Young Open Clusters as Determined
through a Statistical Approach Using 2MASS and GSC Data}

\author{Anandmayee Tej\altaffilmark{1,2,3}, Kailash C. Sahu\altaffilmark{1}, 
T. Chandrasekhar\altaffilmark{2} \& N.M. Ashok\altaffilmark{2}\\
E-mail: tej@newb6.u-strasbg.fr, ksahu@stsci.edu, chandra@prl.ernet.in,
ashok@prl.ernet.in}

\altaffiltext{1}{Space Telescope Science Institute, 3700 San Martin
Drive, Baltimore, MD 21218, USA} \altaffiltext{2}{Physical Research
Laboratory, Navrangpura, Ahmedabad - 380009, India} \altaffiltext{3}
{Present Affiliation: Observatoire Astronomique de Strasbourg, 67000 Strasbourg,
France}

\begin{abstract} 

In this paper we present the mass functions in the substellar regime of
three young open clusters, IC 348, $\sigma$ Orionis and Pleiades, as
derived using the data from the 2 Micron All Sky Survey (2MASS)
catalogue which has a limiting magnitude of $K_{s} \sim 15$, and the
latest version of the Guide Star Catalogue (GSC) which has a limiting
magnitude of \setcounter{footnote}{3} $F\footnote{F refers to the POSS
II IIIa-F passband} \sim$ 21. Based on recent evolutionary models for
low mass stars, we have formulated the selection criteria for stars
with masses below $0.5 M_{\odot}$. Using a statistical approach to
correct for the background contamination, we derive the mass function
of objects with masses ranging from 0.5$ M_{\odot}$ down to the
substellar domain, well below the Hydrogen Burning Mass Limit.  The
lowest mass bins in our analysis are 0.025, 0.045 and 0.055 M$_\odot$
for IC 348, $\sigma$ Orionis and Pleiades, respectively. The resultant
slopes of the mass functions are 0.8 $\pm$ 0.2 for IC 348,  1.2 $\pm$
0.2 for $\sigma$ Orionis and 0.5 $\pm$ 0.2 for Pleiades, which are
consistent with the previous results. The contribution of objects below
0.5 M$_\odot$ to the total mass of the cluster is $\sim$40\%, and the
contribution of objects below 0.08 M$_\odot$  to the total mass is
$\sim$4\%. 

\keywords{stars: low-mass, brown dwarfs; open clusters and
associations: IC 348, Pleiades, $\sigma$ Orionis}

\end{abstract} 

\section{Introduction} 

The Initial Mass Function (IMF) of stars is one of the most fundamental
and crucial ingredients in models of galaxy formation and stellar
evolution. It determines several key parameters in stellar populations,
such as the yield of heavy elements, the mass-to-light ratio,
luminosity evolution over time, and the energy input into the
interstellar medium. The determination of the IMF is therefore of great
astrophysical importance. The IMF of low-mass stars  is of special
interest in this context, since they contain a major  fraction of the
stellar mass, and have been hypothesized to contain a significant 
fraction of  the total mass in the Universe (see, e.g., Fukugita, Hogan
and Peebles, 1998).  In this paper we mainly deal with objects having
masses less than  0.5$M_{\odot}$, the lowest mass of the detectable
objects being as low as 0.025$M_{\odot}$. These low-mass stars evolve
little over the lifetime of the Universe, and hence the observed 
present day Mass Function of these stars is likely to be a good
representation of their IMF. But, the IMF at or below the HBML remains
poorly known,  mainly for two reasons. First, such objects are faint
and hence difficult to detect. Second, the mass-luminosity relation of
these objects is uncertain and model-dependent, and hence their mass
determination is imprecise. Significant improvements are being made on
both these aspects, as described below.

The difficulty caused by their faintness can be greatly alleviated by
concentrating on young, low-mass objects since the low-mass stars at or
below the HBML are expected to be warmer and more luminous when young,
although they rapidly cool and fade with age (Burrows et al. 1997,
D'Antona \& Mazzitelli 1997, Baraffe et al. 1998). Hence young and
nearby open clusters provide a good opportunity to study the low end of
the stellar IMF, since the suitable combination of their youth and
proximity makes it possible to detect objects well below the  HBML in
these clusters, particularly at near-infrared wavelengths.

The uncertainty in the mass-luminosity relation has been greatly
reduced by the tremendous progress in the theoretical models for the
evolution of these cool and dense objects over the past few years.
These models play a crucial role in predicting masses of these low-mass
objects from the observable quantities like colors and luminosities. 
Burrows et al. (1997) have generated models of spectra, colors and
evolution of brown dwarfs using nongrey calculations. Their models span
the mass range of 0.3$M_{J}$ to 70$M_{J}$ (where $M_{J}$ refers to the
mass of Jupiter) with effective temperatures varying from $\sim$1300 K
to 100 K. D'Antona \& Mazzitelli (1997) simulate the evolution of
objects in the mass range 20$M_{J} \leq M \leq 1.5 M_{\odot}$. They
describe the star's evolution from the hydrostatic phases of pre-main
sequence contraction to the hydrogen burning main sequence phase
through deuterium and lithium burning . Baraffe et al.(1998) have
developed the evolutionary models in the 0.075 $M_{\odot}$ to 1
$M_{\odot}$ mass range for solar type metallicities based on the
NextGen atmospheric models of Allard et al. (1996), and Chabrier et al.
(2000) have extended this study by including dust formation and
opacity. The recent models for dwarfs by Marley et al (2002)  take the
extra effect of sedimentation into account, which suggest that some of
the colors, particularly the Sloan i$'$-z$'$, can be greatly affected by
sedimentation.

The advent of the red-sensitive CCDs and 2-dimensional near-IR
detectors in the last decade has made it possible to detect such
low-mass objects, and there have been numerous imaging surveys
targeted towards open clusters to probe the substellar domain. Surveys
by Wilking et al. (1999) and Luhman et al. (1999) for $\rho$ Ophiuchi,
Herbig (1998) and Luhman (1999) for IC 348, B\'{e}jar et al. (1999) and
Zapatero Osorio et al. (1999a) for $\sigma$ Orionis, Zapatero Osorio et
al. (1996) and Stauffer et al. (1999) for $\alpha$ Persei, Zapatero
Osorio et al. (1997;1999b), Bouvier et al. (1998) and Hambly et al.
(1999) for Pleiades, Hambly et al. (1995), Pinfield et al. (1997) and
Magazz\`{u} et al.(1998) for Praesepe, Gizis et al. (1999) and Reid \&
Hawley (1999) for Hyades, and Barrado y Navascu\'{e}s et al. (2001a) for
IC 2391 are to name a few. 

The recent release of the 2MASS catalogue in the near-infrared
wavelengths with a limiting magnitude of $K_{s} \sim 15$, and the
latest (development) version of the Guide Star Catalogue (GSC) with a
limiting magnitude of $F \sim 21$ form an ideal combination to study
low-mass objects in nearby open clusters using a statistical approach. 
We have used these two datasets in conjunction with the recent
evolutionary models to isolate the low mass members of the clusters.
The cluster membership is not ascertained by follow-up spectroscopy or
proper motion studies. But the background/foreground contamination is
accounted for statistically by studying nearby control fields.  In \S
2, we review some previous work on the derivation of the mass function;
 in \S 3, we describe the rationale of our sample selection and the
details of the 3 individual clusters selected for this study; in \S 4,
we describe the data, the procedure  adapted in selecting the cluster
members and the method used for their mass determination; in \S 5, we
describe the specific selection criteria used for the  individual
clusters and the slope of the resultant mass function; and  we end with
a discussion of the results in \S 6. We plan to extend this work to
more clusters in the future.
 
\section{The Mass Function}

The stellar mass function is defined as the number density of stars per 
unit mass bin, and is universally represented as

\begin{equation} \Psi(M) = \frac{dN}{dM} \;\;\;\;\;\;\; stars\; 
pc^{-3}\; M_{\odot}^{-1} \end{equation} 
The stellar IMF for the high-mass stars ($M > 1 M_\odot$) has been long
established and well studied following the pioneering work of Salpeter
(1955). Salpeter (1955) derived the IMF from the luminosity function of
the present day field stars assuming a constant rate of star formation
and correcting for the stellar evolution. In linear units the Salpeter
mass function is given by 

\begin{equation} \Psi(M) \propto M^{-\alpha}
\;\;\;\;\;\;\; stars\; pc^{-3}\; M_{\odot}^{-1} \end{equation} 

\noindent where $\alpha$ = 2.35 for stars in the mass range 1--10
$M_{\odot}$.  The steep slope of the IMF indicates that the low-mass
stars greatly outnumber their high-mass counterparts and account for
the major fraction of the stellar mass.  Miller \& Scalo (1979) and
Scalo (1986) rederived the stellar IMF by extending the study to the
subsolar domain. Their derived values of $\alpha$ are 1.4, 2.5 and 3.3
for the mass ranges 0.1$M_{\odot} \leq M \leq 1.0M_{\odot}$,
1.0$M_{\odot} \leq M \leq 10M_{\odot}$ and 10$M_{\odot} \leq M$,
respectively. 

There have been several studies recently to extend the stellar mass
function to lower masses. Kroupa et al (1993) took the age and
metallicity dependence of the mass-luminosity relation into account in
determining the mass functions. Their derived values of $\alpha$ are
2.7 for stars more massive than 1 M$_\odot$, 2.2 in the mass range 
$0.5 - 1 M_\odot$ and $0.7 < \alpha < 1.85$ in the range $0.08 -  0.5
M_\odot$.  Chabrier (2001) has derived the stellar mass function of the
Galactic disk stars down to the vicinity of the HBML using their
luminosities determined from parallax measurements.  His study
indicates that the mass function may be slightly better represented by
a log-normal or exponential form than the power-law form given by Eq.
1. His results suggest that the mass function  flattens out below 1
M$_\odot$ but keeps rising down to the bottom of the main sequence
which is consistent with the results of Kroupa et al (1993). 

All these studies suggest a flattening of the IMF at the low mass end.
However, the nature of the mass function below the HBML, and
particularly its spatial variation if any, is still poorly known. On
the other hand, to understand the Galactic structure, it is important
to know the temporal and spatial variation of the IMF. 

According to the study by Adams \& Fatuzzo (1996), the IMF  shows a low
mass turn over and they attribute this turn over to the suppression of
the very formation of low-mass objects by winds. Models by Price \&
Podsiadlowski (1995) suggest that the stellar IMF of a cluster should
depend on the cluster density because of the stellar interactions
during the process of accretion. Whereas, Larson (1992) links the
dependence of the stellar IMF on the geometrical structure of star
forming clouds and in particular, he attributes the power law form of
the upper IMF to the presence of hierarchical or fractal structure in
these clouds. The fragmentation of these filaments is predicted to
yield a minimum stellar mass of $\sim 0.1 M_{\odot}$.

The searches for low-mass objects in young clusters by various groups
mentioned in \S 1 have also found many hundreds of candidates with
estimated masses  below the hydrogen-burning limit of $\sim 0.075
M_\odot$. In addition, some of these searches have also discovered
free-floating objects with inferred masses possibly below the
deuterium-burning limit of $\sim 0.013 M_\odot$ (see, e.g. Lucas \&
Roche 2000; Zapatero Osorio et al. 2000). The scenarios proposed for
the formation of such low-mass objects generally involve ejections from
a multiple system. Reipurth \& Clarke (2001) argue that brown dwarfs
are not formed as a result of low mass core collapse but they are
ejected  stellar embryos from multiple systems for which the star
formation process was aborted before the onset of hydrogen burning.
This ejection model could appreciably explain the flattening of the IMF
at low masses. Boss (2001) has suggested that, under certain
conditions, the same collapse and fragmentation process that produce
single and multiple stars can also produce planetary-mass,
self-gravitating objects. The ejection of these fragments from an
unstable, protostellar system may explain the formation of the
free-floating planetary-mass objects found in some young clusters. 

{\begin{figure}
\plotone{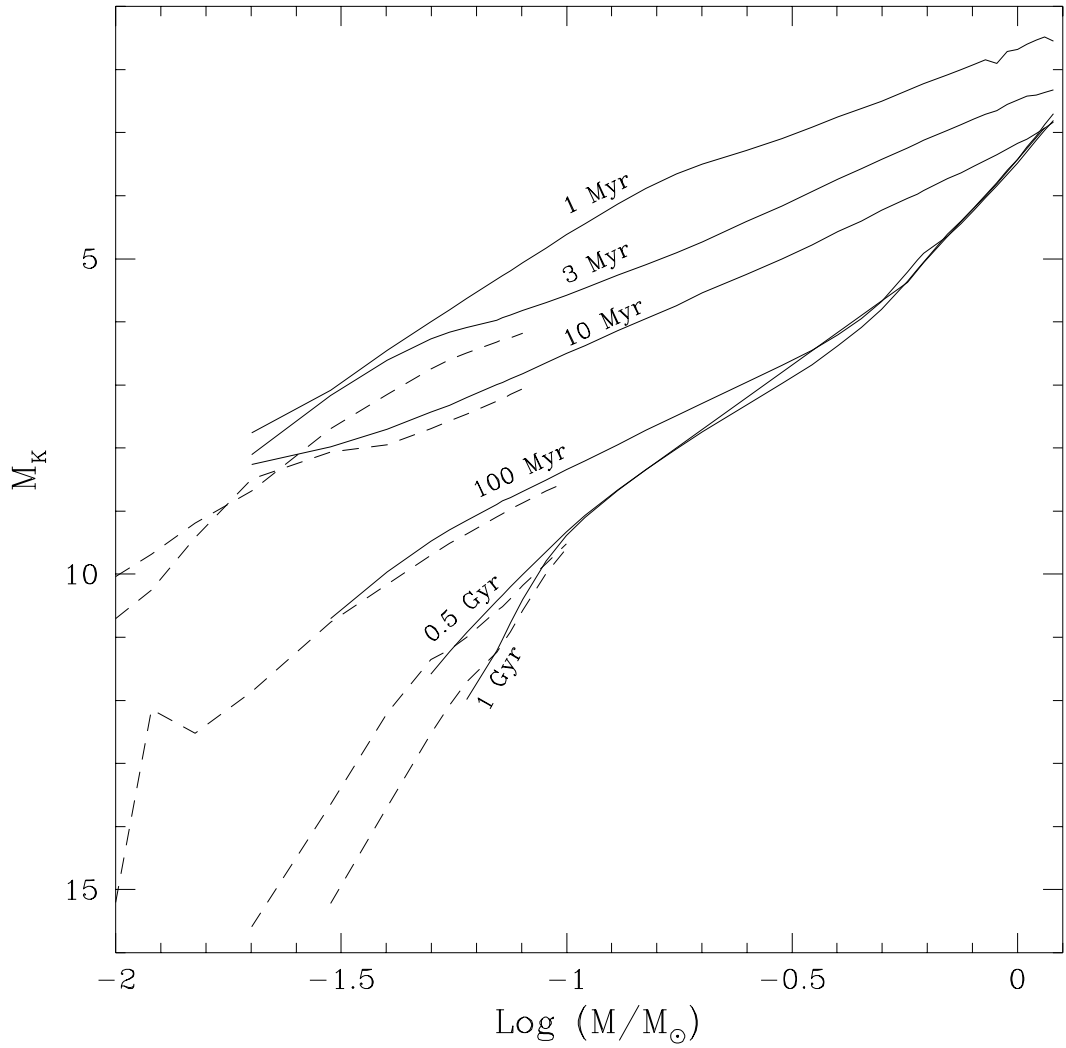}
\figcaption{In this figure we show the absolute $K$ magnitudes
as a function of mass (in logarithmic units) for different ages. The
isochrones are from the models by Baraffe et al. (1998) (solid lines)
and the Dusty models by Chabrier et al. (2000) (dashed lines).}
\end{figure}}

Reid et al. (1999) derive the mass function of the field stars using
the 2MASS and the DENIS data within 8 pc of the Sun in the mass range
0.1 -- 1 $M_{\odot}$. Their study shows that in the lower mass range
the mass function is flatter ($1 \leq \alpha \leq 2$), the value of
$\alpha$ being closer to the lower limit. Extrapolating their results
to the substellar regime down to 0.01 $M_{\odot}$ they predict that the
ratio of the number of brown dwarfs to the main sequence stars is 2:1
in the solar neighborhood. Their study implies that brown dwarfs
contribute less than 15\% of the total mass of the disk and hence they
are unlikely to be the major constituent of Galactic dark matter.
However, their mass function is necessarily derived from a mixed sample
of stars formed at different times and different environments.

Studying the stellar IMF in different clusters provides a means to
check whether the IMF in clusters is different from that of the field
population and also to check the universality of the IMF.  Low mass
objects can, in principle, escape a cluster due to cluster dynamics and
Galactic tides. In addition, the internal velocity dispersion may
result in mass segregation in the cluster which is expected to increase
with cluster age (Spitzer \& Mathieu 1980; Kroupa 1995). However,
clusters younger than one relaxation time should not show these
effects. Mass segregation has been observed in very young ($\leq$ 1
Myr) clusters such as the Orion Trapezium (Hillenbrand \& Hartman
1997), but these are most likely the result of the formation process
rather than the dynamical evolution. The typical relaxation time of
clusters like Pleiades is $\sim$ 10 Myr (Raboud \& Mermilliod 1998).
Hence, clusters with ages younger than $\sim$ 10 Myr would be ideal to
understand the mass function at the low mass regime, as these clusters
are not old enough to have lost members due to stellar evolution, or to
have suffered mass segregation due to dynamical effects such as
evaporation or violent relaxation (Lada \& Lada 1991 and references
therein). Moreover, young clusters are ideally suited for the detection
of very low mass objects and brown dwarfs and hence for deriving the
mass function in the substellar domain. This fact can be better
appreciated in Figure 1 which shows the $K$ magnitudes as a function of
age and mass of the low-mass objects. The figure, which uses the models
\setcounter{footnote}{0} of Baraffe et al. (1998)\footnote{The model
isochrones in the published version constitute a subset of the
available models and go down to 0.075$M_{\odot}$. A larger set of model
isochrones spanning a more extensive grid of ages and masses are
available electronically via anonymous ftp from {\it ftp.ens-lyon.fr}.}
and Chabrier et al. (2000), shows that objects with masses less than
0.1 M$_\odot$ are more than 5 magnitudes brighter when they are 1 Myr
old than when they are 1 Gyr old, which makes them more easily
detectable.

\section{Sample Selection}

The main objective of the present study is to derive the stellar mass
function down to very low masses for a sample of young open clusters,
in an attempt to obtain a global view of the mass function of low-mass
objects in such young clusters. Low mass stars ($M \leq 0.5M_{\odot}$)
have effective temperatures below 3500 K (Berriman \& Reid 1987) and
brown dwarfs ($M \leq 0.08 M_{\odot}$) have temperatures below 2800 K
(Chabrier et al. 2000), which implies that the spectral energy
distribution of these objects peaks in the near infrared between 1 -- 3
$\mu$m. Hence, the near IR bands are ideal for detecting these objects.
We use the data from the 2MASS Second Incremental Release and the
latest version of the GSC catalogue. This 2MASS release covers $\sim $
47\% of the sky in $JHK_{s}$ near-IR bands. Fortunately, it covers the
major area of the three clusters and their corresponding control
fields chosen by us. The incomplete coverage of our fields by the
2MASS data is accounted for by appropriate normalization as explained
in \S 5. The latest version of the GSC catalog is an all-sky catalog
of stars in the $IIIa -J$ and $F$ bands with limiting magnitudes
of 22 and 21 in each passband respectively. (Currently, this catalog 
has restricted access, and is in its final stages of preparations for 
release, expected to be in late 2002, McLean private 
communication\footnote{http://www-gsss.stsci.edu/gsc/gsc2/GSC2home.htm}).
The red plates used in making the GSC
catalogue had used slightly different emulsions for the Northern
and the Southern surveys. Accordingly, the F-passbands 
are slightly different for
the northern and the southern sources, which we have taken into account in
our analysis of the three clusters. The 2MASS
\setcounter{footnote}{2}
catalogue\footnote{Available from
http://www.ipac.caltech.edu/2mass/releases/second/index.html} has roll
over limiting magnitudes of 15 in $K_s$, 15.5 in $H$ and 16.5 in $J$
(the photometric accuracies are shown in Fig. 7 of the documentation
pages of the 2MASS Second Incremental Release). This allows us to probe
low-mass objects down to about 0.03 $M_\odot$ of all nearby clusters (D
$<$ 200pc) with ages less than about 100 Myr (see Fig. 2).

{\begin{figure}
\plotone{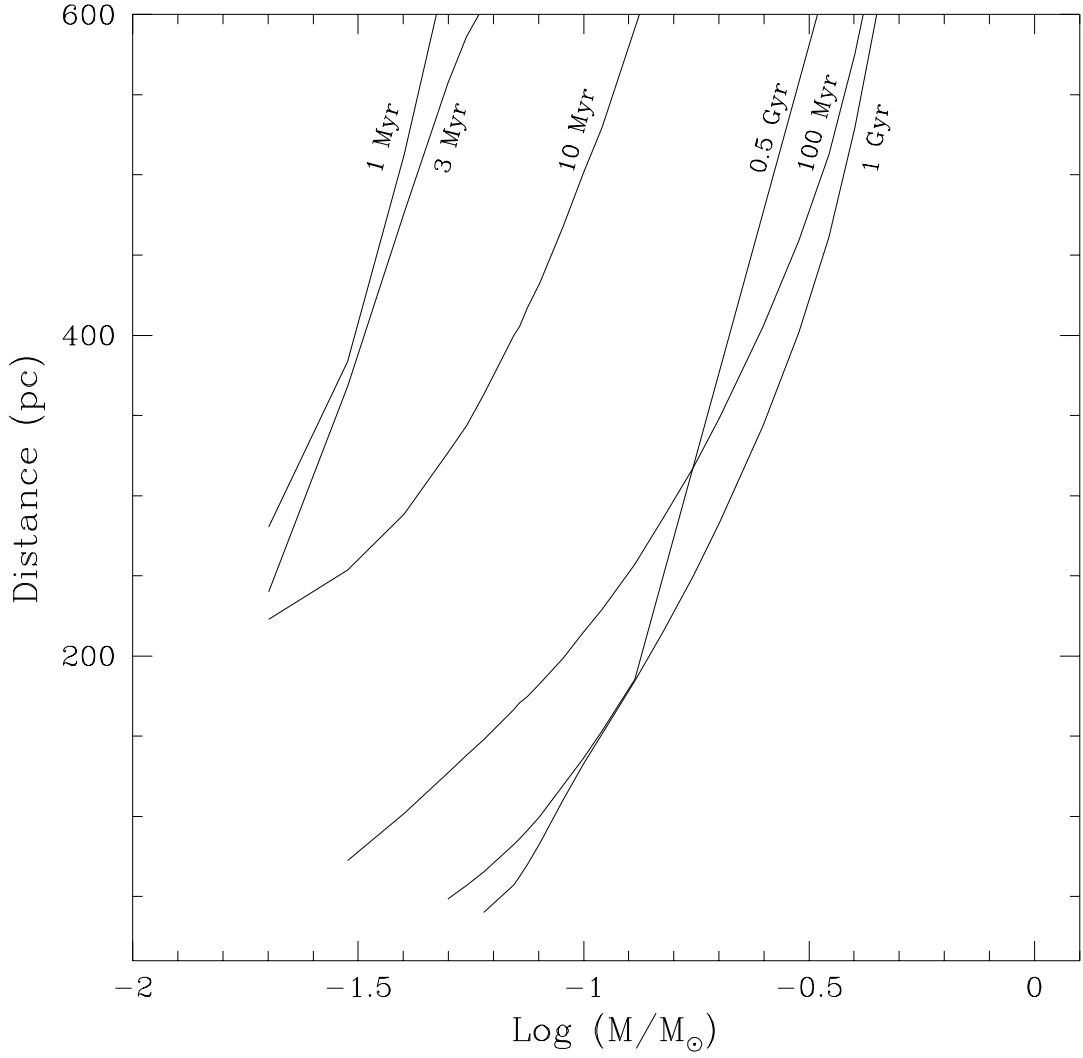}
\figcaption{This figure shows the mass (in logarithmic units)
as a function of distance for different isochrones derived for a
limiting magnitude of 15 in the $K$ band. The model isochrones are from
Baraffe et al. (1998).}
\end{figure}}
Unlike most other previous studies which rely on confirming the
candidate low-mass objects through spectroscopic observations, we use a
statististical approach to estimate the number of low-mass objects. In
a statistical approach, it is important to use several control fields
close to each cluster to subtract the contribution of foreground and
background objects. The nature of these two extended surveys enables us
to use several such control fields and we are not limited by our choice
of the field sizes for the clusters or the control fields. To establish
the viability of a statistical approach, it is important to apply this
procedure for a few well-studied clusters. The clusters we have chosen
for this work are Pleiades, IC 348 and $\sigma$ Orionis, all of which
are young and nearby. The masses of the objects  corresponding to the
faint limiting magnitudes in each passband for the 3 clusters are given
in Table 1. For IC 348 and $\sigma$ Orionis, a member with mass $\sim$
0.025 $M_{\odot}$ would have a $K_{s}$ magnitude $\sim$ 15 and for the
relatively older cluster Pleiades, $K_{S} \sim$ 15 corresponds to
objects with masses $\sim$ 0.04 $M_{\odot}$. Pleiades and IC 348 have
been well-studied through ground-based optical and near-IR photometry
and optical spectroscopy, and hence it would be useful to check the
consistency of our results with that of the previous studies. Previous
studies of $\sigma$ Orionis are confined to photometry in optical
wavelengths, but our study includes the near IR data. A brief
description of the individual clusters is given below. The details of
the derivation of mass functions and comparison with other studies will
be discussed in a later section.

\subsection{IC 348}

IC 348 is a relatively dense, rich and compact cluster at a distance of
$\sim$310 pc, with a size $\sim$ 20$'$ (Luhman 1999) and located
in one end of the Perseus molecular cloud. Herbig (1998) found a
significant spread in the age of the cluster (0.7 -- 12 Myr) based on
imaging observations in $BVRI$. Ground based $JHK$ imaging by Lada \& Lada (1995)
revealed a central subcluster with a radius of 0.5 pc containing half
the cluster members. Luhman et al. (1998) derived an age spread of 5 --
10 Myr for this subcluster which was confirmed by Najita
et al. (2000). There have been several surveys to detect and catalog
the low mass members of this cluster (Luhman et al. 1998; Luhman 1999;
Najita et al. 2000). Luhman et al. (1998) obtained infrared ($K$ band)
and optical spectroscopy and $JHK$ photometry of the stellar population
within the $5' \times 5'$ core of this cluster to study the star
formation, disk properties and the mass function. In a continuing
program to identify and characterize the low mass stellar and
substellar populations in this cluster, Luhman (1999) carried out a
wide and deep photometric survey in $R$ and $I$ covering a total area
of $25' \times 25'$. Low resolution optical spectroscopy of a subset of
the candidate substellar objects found from the survey was performed to
confirm their membership. In conjunction with the evolutionary models
of D'Antona \& Mazzitelli (1997) and Baraffe et al. (1998), Luhman
(1999) concluded that the HBML for this cluster occurs at a spectral
type of M6 and that several objects found in the survey fall below the
substellar boundary with masses as low as 20-30 $M_{J}$. Najita et al.
(2000), using the $HST$/NICMOS narrow band imaging, investigated the
low mass population down to the deuterium burning limit. Their study
spans a mass range of $\sim$ 0.7 $M_{\odot}$ to 0.015 $M_{\odot}$. 

\subsection{$\sigma$ Orionis}

The Orion complex is one of the richest star forming regions in our
Galaxy. $ROSAT$ observations in the Orion belt led to the discovery of
a large sample of X-ray sources near the bright young multiple star
$\sigma$ Orionis (Walter et al. 1994) which belongs to the Orion 1b
association. Subsequent photometry and spectroscopy of the X-ray
sources detected by $ROSAT$ revealed the existence of a young stellar
cluster (Wolk 1996). This cluster is located at a distance of 352 pc.
The age of this cluster is in the range of 1 -- 5 Myr (B\'{e}jar et al.
1999 and references therein) and its estimated size is 25$'$ (Lynga
1983). B\'{e}jar et al. (1999) present the CCD observations in the $R$,
$I$ and $Z$ bands covering a total area of 870 $arcmin^{2}$ around
$\sigma$ Orionis, which led to the detection of objects with masses
down to $\sim$0.02 $M_{\odot}$. By combining results from imaging
surveys and follow up photometric and spectroscopic observations,
B\'{e}jar et al. (2001) identified 64 very low mass members in this
cluster, in the mass range 0.2 $M_{\odot}$ to 0.013 $M_{\odot}$.
Barrado y Navascu\'es et al. (2001b) have recently reported
spectroscopic observations of planetary mass candidates in this
cluster.  

\subsection{Pleiades}

Pleiades is by far the best studied open cluster for low-mass stars.
With an age of $\sim$120 Myr and a distance of 125 pc,
it is an ideal hunting ground for low mass stars and brown dwarfs.
Imaging surveys to detect brown dwarfs began almost a decade ago
(Jameson \& Skillen 1989; Stauffer et al. 1989,1994; Simons \& Becklin
1992; Rebolo et al. 1995; Cosburn et al. 1997; Zapatero Osorio et al.
1997,1999b; Bouvier et al. 1998; Festin 1998). Zapatero Osorio et al.
(1997) reported the $JHK'$ observations of some of the least luminous
members of Pleiades and proposed a substellar limit at spectral type
M7. In a deep $IZ$ survey covering an area $\sim$ 1 $deg^{2}$ in the
central region, Zapatero Osorio et al. (1999b) detected substellar
candidates ranging in mass from 0.075 $M_{\odot}$ to 0.03$M_{\odot}$.
Bouvier et al. (1998) performed a wide field imaging survey of the
Pleiades cluster in the $R$ and $I$ passbands covering a large area of
$\sim$ 2.5 $deg^{2}$ to a completeness limit of $R \sim$ 23 and $I
\sim$ 22. They found 17 objects satisfying the criteria for brown dwarfs
ranging in mass from the HBML down to 0.045$M_{\odot}$. 

\section{The methodology} 

\subsection{The data} 

As mentioned earlier, we have used the data from the 2MASS sky survey
and the recently made available Guide Star Catalogue in this study. In
selecting the 2MASS sources, we have constrained our sample to include
only those sources for which the error in the $K_{s}$ band is less than
or equal to 0.15 mag to avoid the noisy data at the faint end. This
would translate to a signal-to-noise ratio of $\geq$7 in $K_{s}$.
This limit ensures positive detection in the other two bands as well. 
We have also taken into account the different flags so that our sample
is not affected by the spikes of nearby bright stars, contamination
from extended sources, or saturation in any of the three bands. Care is
taken to exclude asteroids and minor planets, identified by the 2MASS
catalog. 

We merged the 2MASS data with the GSC sources by taking the 2MASS
coordinates, and cross-correlating them with the GSC catalogue. For
such a correlation, it is important to take an appropriate search
radius which takes into account the uncertainty in the coordinates. 
The uncertainty in the coordinates mainly comes from the frame of
reference used in the two catalogs. Comparison of Hipparcos data with
the GSC data reveals that the GSC coordinates can differ by as much as
$2''$ from the coordinates in the Hipparcos reference frame that 2MASS
uses (Bakos, Sahu and Nemeth, 2002). A search radius of $2''$ was found
to be an optimum value for this cross-correlation, which was small
enough to reject spurious and multiple detections which could be the
result of diffraction spikes due to bright stars in the field, and
large enough to include any positional uncertainties in the two
catalogs. 

\subsection{Initial selection of cluster members}

Assigning cluster membership to sources is a difficult task. The color
magnitude diagrams [CMDs] for various clusters given in the literature show
that the density of faint members is usually comparable or even
significantly less than the field/background/foreground population
(Barrado y Navascu\'{e}s et al.(2001a) for IC 2391, Barrado y
Navascu\'{e}s et al.(2001c) for M35 etc.) The use of CMDs can be used
as an effective tool to delineate possible members from the field
contaminants. Of course, proper motion and/or spectroscopic studies are
needed to establish the membership of specific sources.

The CMDs best suited for this purpose can be determined by examining
the locations of sources with different masses and temperatures in 
different CMDs, and the slope and direction of the low-mass main
sequence and the reddening vectors. An example $K_{s}$ versus
$(J-K_{s})$  CMD  is shown in Figure 3 where the dots represent the
stars in the direction of the Pleiades cluster. To differentiate the
line-of-sight contaminants from the cluster members we need a reliable
locus for the low-mass main sequence. 

{\begin{figure}
\plotone{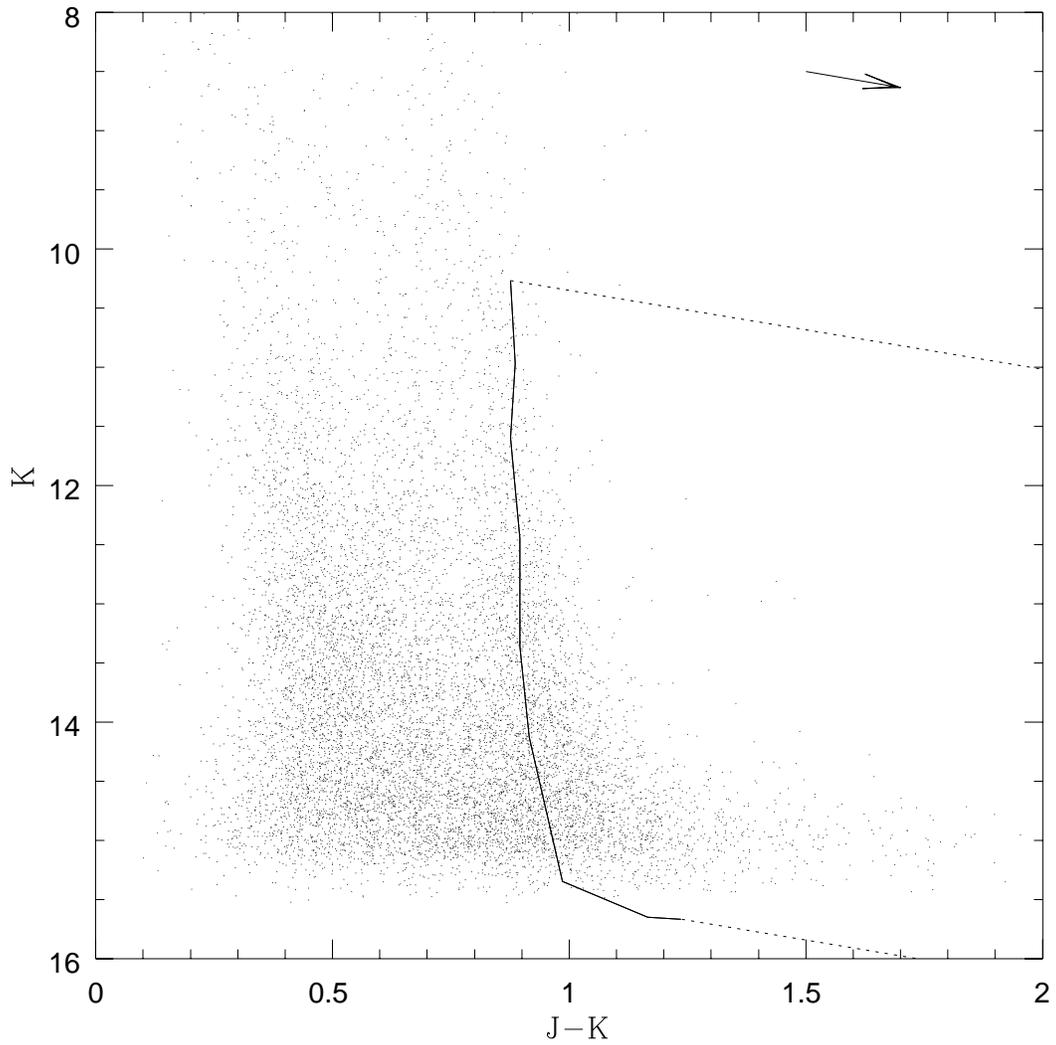}
\figcaption{This figure shows the $K_{s}$ versus $(J-K_{s})$
plot for the Pleiades cluster. The Leggett main sequence is shown as
the solid line from M0 and beyond. The dotted lines are the redenning
vectors as derived from Rieke and Lebofsky (1985). The arrow indicates
the direction of redenning. }
\end{figure}}
These open clusters are located in the spiral arm of the Galaxy which
comprises of the young disk population.   
To account for the contaminants, we construct a $K_s \sim (J-K_s)$
locus for the young-disk objects using the empirical data from Table 6
of Leggett (1992). The solid line in Fig. 3 shows such a disk
sequence for all stars with spectral types M0 (corresponding to 0.5
M$_\odot$) and later as taken from Leggett (1992). Here, we have
appropriately shifted  the Leggett sequence to take the distance and
the extinction of the Pleiades cluster into account. The lowest two
points in this line are identified as {\it young-old} (Y/O)  by Leggett
(1992). We derive the reddening vectors from the interstellar
extinction laws of Rieke \& Lebofsky (1980) and Bessell \& Brett
(1988); these reddening vectors are shown as dashed lines in the Fig.
3.  We have chosen to plot the young-disk population because such a
plot is likely to be most useful in rejecting the  
older non-member stars along the line-of-sight which are expected to
fall to the left of this sequence owing to metallicity effects. 

As seen from the figure, the stars with spectral types M0 and later
fall more or less vertical in this diagram, and the $(J-K)$ color
changes by $\sim$ 0.35 over this entire spectral range. The reddening
vector runs almost perpendicular to this sequence with a slope of
$\sim$0.67. Since the effect of the distance is to move a particular
source vertically in this diagram, there is a clear degeneracy between
mass and distance for the mass range from 0.6 $M_{\odot}$ to $\sim$
0.06$M_{\odot}$ (Leggett 1992). For lower masses, the extinction vector
runs almost parallel to the direction of the Leggett sequence
creating a degeneracy between mass and extinction in this spectral
range. 

The inclusion of one optical color in the CMD [e.g. $F$ versus $(F-J)$]
overcomes the aforementioned degeneracies as explained below. One
problem, however, is that the original data of Leggett (1992) give the
absolute magnitudes and the intrinsic colors for the standard broad
band filters in the Cousins, Johnsons and the CIT systems, and  not in
the F-passband. So we need to first convert the R magnitudes to F
magnitudes by applying appropriate $(F-R)$ color corrections. For this
purpose, we compared the Leggett young disk sequence with model
isochrones of different ages in a $F$ versus $(F-R)$ plot and found
that the 100 Myr reasonably fits the Legett young disk sequence as
expected (e.g. Allen 1973). So we used the 100 Myr model isochrones in
F and R to calculate the values of $(F-R)$, and applied this correction
to the empirical $R$ values derived from Leggett (1992). The resulting
CMD in $F$ versus $(F-R)$ is shown in Fig.4, where the color increases
rapidly with spectral type ($\sim$ 3 mags) and the reddening vector is
steeper with a slope of $\sim$1.59.  Use of this plot avoids the
degeneracies found in the  $K_s \sim (J-K_s)$ CMD by minimizing the
overlap between the reddened background stars and the low mass members
of the cluster, and offers an efficient rejection criterion for
non-members. Hence we have used the $F$ vs. $(F-J)$ CMD to delineate
field stars from the cluster members. Figure 4 shows a clear separation
between the cluster and non-cluster members, which justifies the use of
this CMD.

{\begin{figure}
\plotone{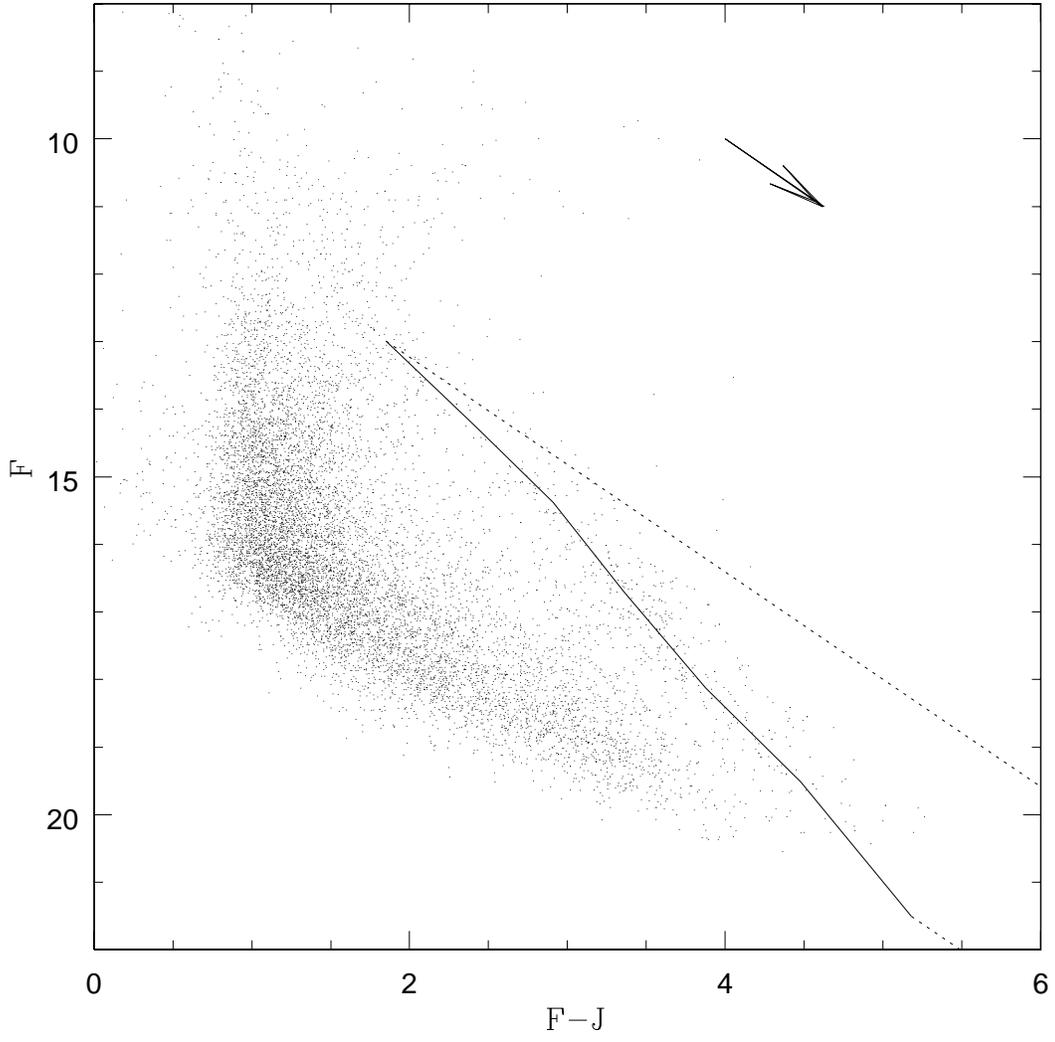}
\figcaption{Same as Fig. 3 showing the $F$ versus the $(F-J)$ plot.}
\end{figure}}

While rejecting the field star population, care has to be taken not to
reject genuine candidates from the sample. The disk population in the
spiral arms are the field contaminants in the line-of-sight. As will be
shown later, the main sequence from Leggett (1992), corresponding to
this young disk population, falls to the left of the theoretical
isochrones from Baraffe et al. (1998) for the clusters IC 348 and
$\sigma$ Orionis. Hence, this sequence from Leggett (1992) is preferred
over the other for the rejection criterion for these two clusters. For
the Pleiades cluster, which is $\sim$100Myr old, the two tracks overlap
for most of the masses but the Baraffe model falls to the left for
some part of the low mass sequence and hence is preferred over the
Leggett main sequence to reject non members, and at the same time to
ensure that none of the genuine members are rejected. Note that
throughout the selection process, we have erred on the side of
including some contaminants while ensuring that we do not reject {\it
any} genuine low-mass objects. The price we pay in following this
conservative approach is that our sample will inevitably include some
foreground/background contamination. This is, however, a small price to
pay since the contamination can be corrected through observations of
the control fields. On the contrary, if some of the genuine objects are
rejected, the effect can be disastrous since the observed number of
stars would then be smaller than the actual number, thus making the
statistics potentially incorrect. It is worth noting here that, since
we have followed this conservative approach, the sources lying between
the Leggett main sequence and the theoretical isochrone should be
treated with caution.

Even though the $F$ vs. $(F-J)$ CMD is a better choice to separate the
cluster members from the non-members, we still see a smooth transition
between the field stars and the cluster members in Fig. 4 which
implies that our cluster sample would still be contaminated. This
contamination could be due to the background reddened stars as well as
the foreground population. We need to correct for this contamination
before we can derive a reliable mass function. This becomes especially
important in our study since we are not carrying out follow-up
spectroscopic or proper motion studies to confirm the membership of
individual members. As we will demonstrate in detail later, the
contamination can be satisfactorily taken into account by studying the
properties of off-fields or control fields in the vicinity of the
clusters. We have tried to use several control fields, each covering
the same area as the cluster, which are located near the cluster and
are free from anomalies. These control fields were selected by visually
inspecting the star density from the POSS plates. For IC 348, we have
used the CO maps of Bachiller \& Cernicharo (1986) to find suitable
control fields. Table 2 gives the location and sizes of the cluster
and control fields used in our study.

\subsection{Stellar mass determination}

The masses of our selected candidates are determined by comparing the
observed magnitudes with those predicted by the evolutionary models. 
We have used the evolutionary tracks of Baraffe et al. (1998) for this
purpose, as they provide the magnitudes and colors as a function of
mass for various ages in the passbands of interest. The magnitudes in
the F-bands were specially calculated and were kindly provided to us by
Baraffe and Allard at our request, which are used in this analysis.
Whereas, in order to transform the effective temperatures and
luminosities of other models we need to use the bolometric corrections
of the Baraffe model. Additionally, the tracks by Baraffe et al. (1998)
have been successful in fitting the mass-luminosity relation in various
optical and infrared passbands and predicting coeval ages for members
of several young multiple systems (White et al. 1999; Luhman 1999). It
is also seen that these models provide good fits to the infrared
photometric sequence in the Pleiades and $\sigma$ Orionis clusters
(Mart\'{i}n et al. 2000; Zapatero Osorio et al. 2000). 
It is worth emphasizing here that the mass determinations are model
dependent and are subject to possible systematic effects arising from
the different parameters used in the models.

\section{Low mass members and the mass function} 

In this section we discuss the selection criteria and the resulting
mass functions derived for the individual clusters.

\subsection{IC 348}

To sample the entire cluster, we consider all the sources within a
radius of $20'$ around the cluster center of IC 348. The wide and deep
study of Luhman (1999) also covers almost the entire cluster, and
includes optical photometry and follow-up spectroscopy. The study by
Luhman et al. (1998) includes both optical and near-IR observations
for the central $5'\times 5'$ of the cluster. Since our study includes
both optical and near-IR photometry, IC 348 provides an ideal
opportunity to compare the results, and to check the consistency
between the different approaches. For this cluster, we derive the
value of interstellar reddening from the sample of 70 confirmed low
mass cluster members of Luhman (1999). The observed distribution of
the extinction values listed in Luhman (1999) ranges from $A_{V}=0.0$
to $A_{V} \sim 8$ with a peak around $A_{V}$=0.3 magnitudes. We have
used the lower value of $A_v= 0$ magnitudes in our selection criteria
to ensure that none of the genuine cluster members are rejected. The
fact that some members of IC 348 have zero extinction is also seen in
the data of Najita et al. (2000) where there are a few sources with
$A_{K}=0.0$.

Fig. 5 shows the $F$ versus $(F-J)$ plot for all the sources in this
region of the cluster and the two control fields. The figure also shows
the main sequence from Leggett (1992) and the theoretical isochrone for
5 Myr from Baraffe et al. (1998) appropriately scaled to a distance of
316 pc (Herbig 1998) and an interstellar reddening of $A_{V}$ of
zero.

{\begin{figure}
\plotone{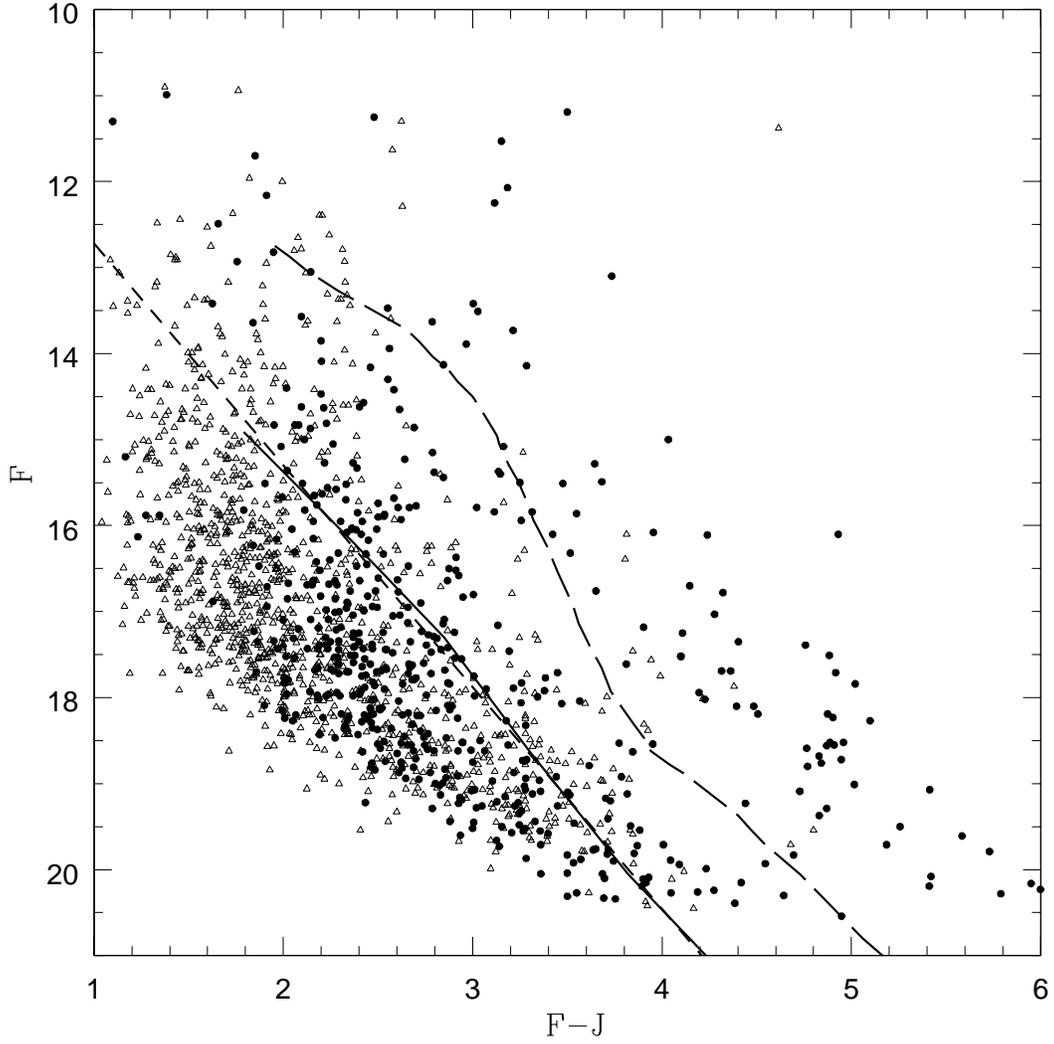}
\figcaption{The $F$ versus the $(F-J)$ plot 
for IC 348 (closed circles). 
Also plotted are the sources from the control fields (open triangles),
the Leggett main sequence (solid), the straight line fit to the
Leggett main sequence (short dashed) and the 5 Myr isochrone from
Baraffe et al. (1998) (long dashed). For clarity we have plotted
every second point of the cluster and the control fields.}
\end{figure}}

The number of sources increases to the left of the main sequence locus
which can be attributed to the field star contamination. Hence, the
objects bluer and fainter than the Leggett main sequence are rejected,
giving us the first criterion for the choice of candidates, which is
derived by fitting a straight line to the data points of Leggett (1992)
\begin{equation} F < 2.58(F-J) + 10.14 \end{equation}

We then use the model isochrones from Baraffe et al.(1998) in other
CMDs appropriately scaled to the distance and extinction of IC 348, to
select the low mass members of the cluster. As discussed in section
3.1, this cluster exhibits appreciable spread in the age. We adopt a
mean age of 5 Myr and use the magnitudes and colors of the Baraffe
models to formulate the following criteria for selecting the objects
with mass $ < 0.5 M_{\odot}$.

\begin{equation} F - K \geq 4.09 \end{equation} 
\begin{equation} K \geq 11.20 \end{equation} 
\begin{equation} J - K \geq 0.90 \end{equation}
\begin{equation} H - K \geq 0.20. \end{equation}

{\begin{figure}
\plotone{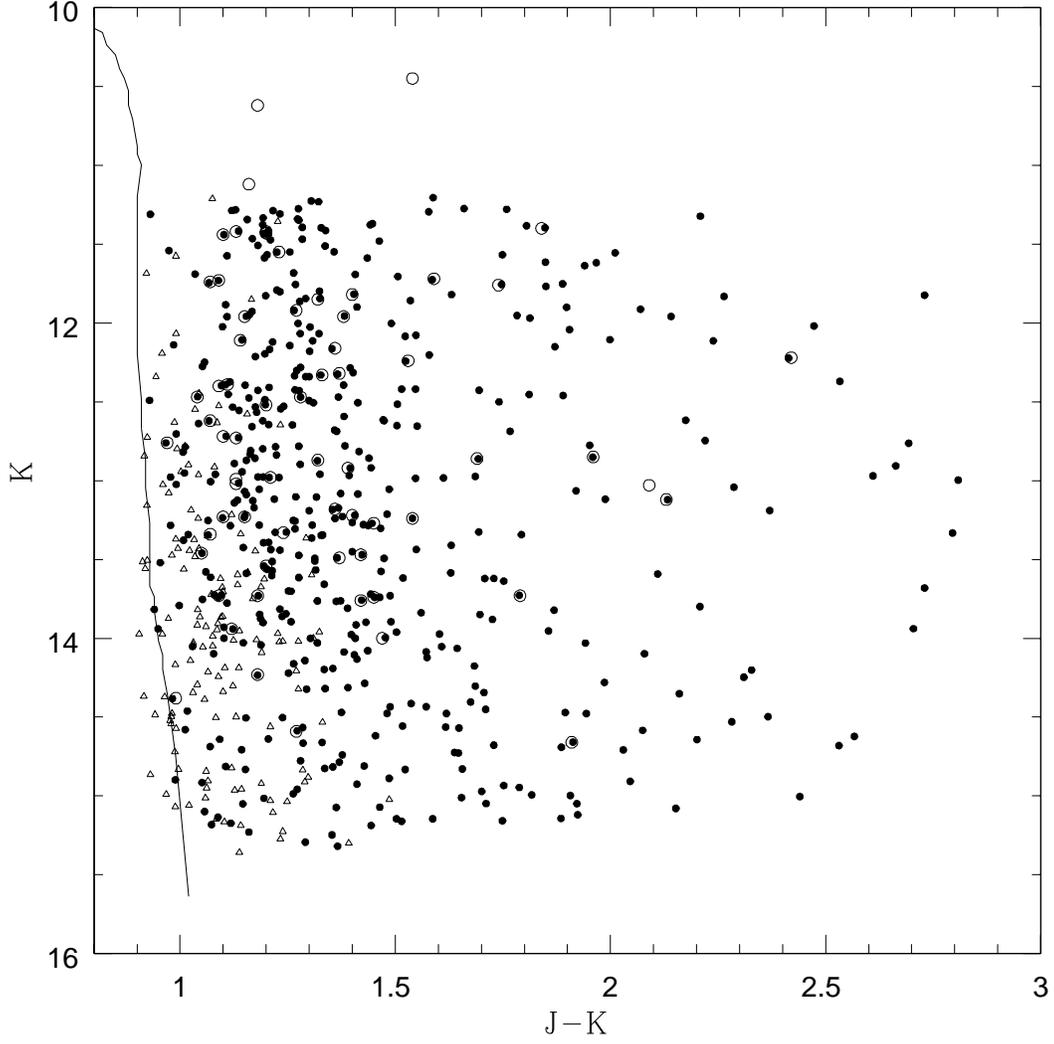}
\figcaption{The $K_{s}$ versus $(J-K_{s})$ diagram for the
sources satisfying the low mass criteria, IC 348 (closed circles) and
control fields (open triangles). Also plotted is the theoretical
isochrone of 5 Myr from Baraffe et al. (1998)(solid line). This plot
also shows the position of the confirmed low-mass members of this
cluster from Luhman (1999)(open circles).}
\end{figure}}

The sources satisfying all the four criteria are designated as
potential low-mass members of the cluster and the sources in the
control fields which satisfy these criteria are used to correct for the
possible contaminants. Fig. 6 shows the $K_{s}$ versus $(J-K_{s})$
plot of the sources satisfying the low mass selection criteria. The
closed circles represent the sources from the cluster and the open
triangles are the candidates from the control fields which represent
the possible contaminants. The plot also shows the theoretical
isochrone for a 5 Myr cluster scaled to the distance and reddening of
IC 348. As one can see, there are a few sources from the control
samples to the right of the theoretical isochrone. We correct for this
contamination while deriving the mass function. Also plotted are the
confirmed low mass members from the Luhman (1999) sample (open
circles). The fact that they all fall in the region satisfying all our
selection criteria proves the validity of this approach. This figure
also suggests low internal extinction in this cluster. If the internal
extinction were high, the stars in the near side of the cluster would
undergo less extinction, and the stars on the far side would undergo
more extinction. As a result, the observed stars would be expected to
fall on both sides of this theoretical main sequence. Thus, the fact
that very few stars lie to the left of the main sequence implies that
most of the extinction must be foreground rather than internal. On the
other hand, we do observe many stars on the right side of the
theoretical sequence. Since the internal extinction is not large, this
would argue that the most of the observed extinction is not
interstellar, but circumstellar. In a recent work Muench et al. (2001)
find, from the analysis of the $JHK$ colors, that $\sim$50 \% of the
brown dwarf candidates in the Trapezium cluster display significant
infrared excess. This suggests that these sources are extremely young
and provides independent confirmation of their cluster membership and
low mass nature. Fig. 7 shows the $(J-H)$ versus $(H-K)$ color - color
diagram for the low mass stars of the cluster and the control fields.
In this figure, the region to the left of the redenning band is
forbidden for young stellar objects. Hence, the location of sources in
this region is most likely due to the combined uncertainties in the
models, the derived magnitudes and the extinction characteristics. The
region to the right of the redenning band is occupied by sources with
infrared excess (Lada \& Lada 1995). This is in agreement with the
discussion on the circumstellar extinction.

{\begin{figure}
\plotone{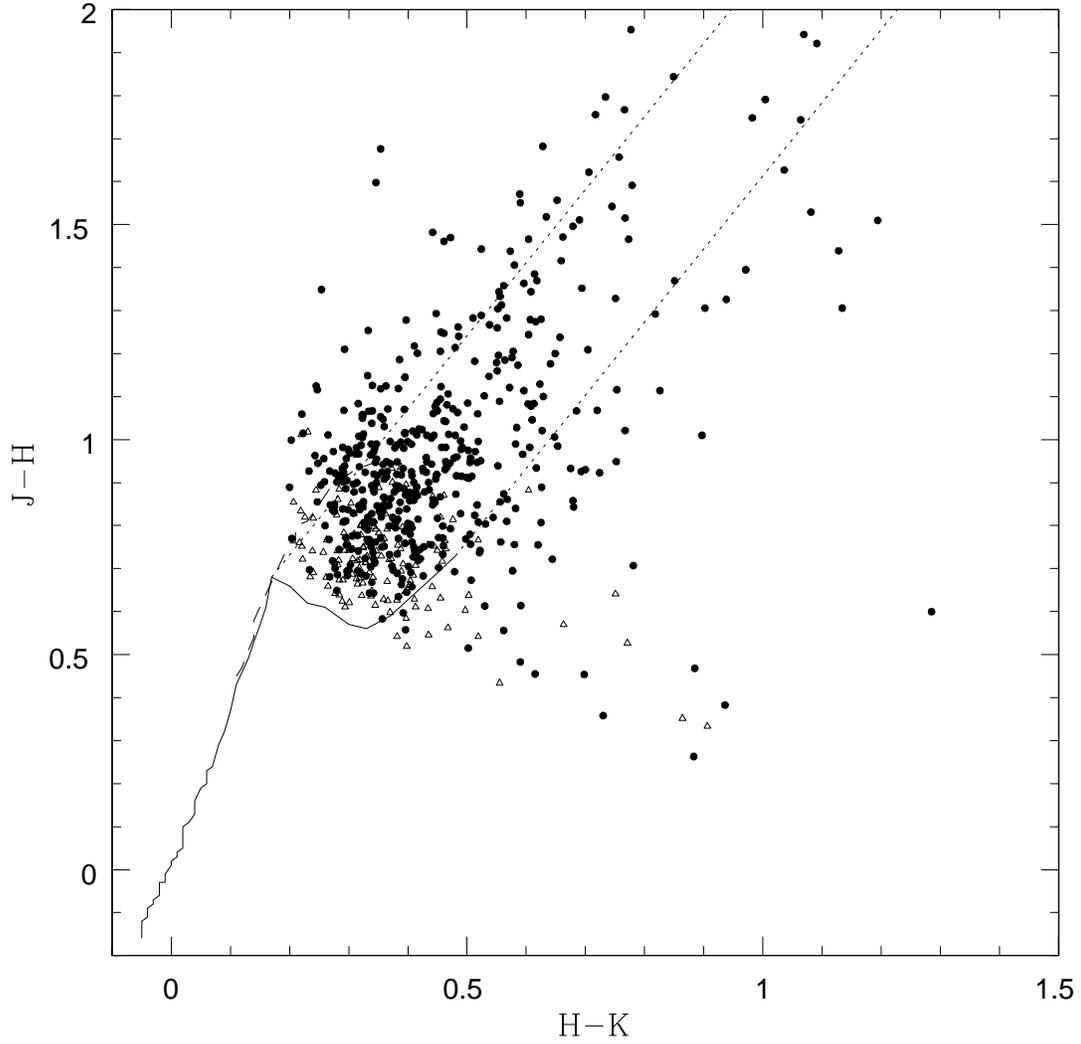}
\figcaption{The $(J-H)$ versus the $(H-K_{s})$ color -- color
plot for IC 348 (closed circles) and the control fields (open triangles).
The solid and the dashed lines are the main sequence and the giant
sequence from Koorneef (1993). The dotted lines are the reddening
vectors derived from Rieke \& Lebofsky (1980).}
\end{figure}}

To derive the mass function, we construct mass bins of 0.01 $M_{\odot}$
in the mass range 0.02 -- 0.09 $M_{\odot}$. As the number of candidates
decreases for higher masses, the width of the bin is increased to 0.1
$M_{\odot}$ for the mass range 0.1 -- 0.5 $M_{\odot}$.
The members of the cluster field and the two control fields satisfying
the five selection criteria are grouped into these mass bins. The
grouping is done based on the model-predicted $K$ magnitudes for these
mass bins taking the distance, age and mean extinction of IC 348 into
account. The number of sources in each mass bin is then normalized with
the total number of sources detected in that field.  
This is essential in order to correct for the unequal area covered by
the cluster and the control fields, often caused by incompleteness of
the survey. The mean percentage of the number of sources present in
each mass bin of the control fields quantifies the contamination one
expects. This percentage contamination is then removed from the cluster
bins and divided by the bin width to obtain the value of $dN/dM$. This
as a function of the mass gives the mass function (Eqn 1 \& 2).  Fig.
8 shows the resultant mass function for IC 348. The closed circles are
based on the  models of Baraffe et al. (1998) and the open triangles
are derived using the models of Chabrier et al. (2000) which include
dust opacities. The error bars in $y$ show the $\sqrt{N}$ errors
involved in the counting statistics. The least squares fit to the data
points derived from the Baraffe et al. (1998) model yield a slope of
$-0.7 (\alpha = 0.7$), with an uncertainty of $\pm 0.2$. In deriving
this slope, we have rejected the lowest mass bin since this is affected
by the limiting magnitude of the 2MASS survey.  As seen from this
figure the results from the models of Chabrier et al. (2000) are in
agreement with the non-dusty models of Baraffe et al. (1998) within the
given scatter.

{\begin{figure}
\plotone{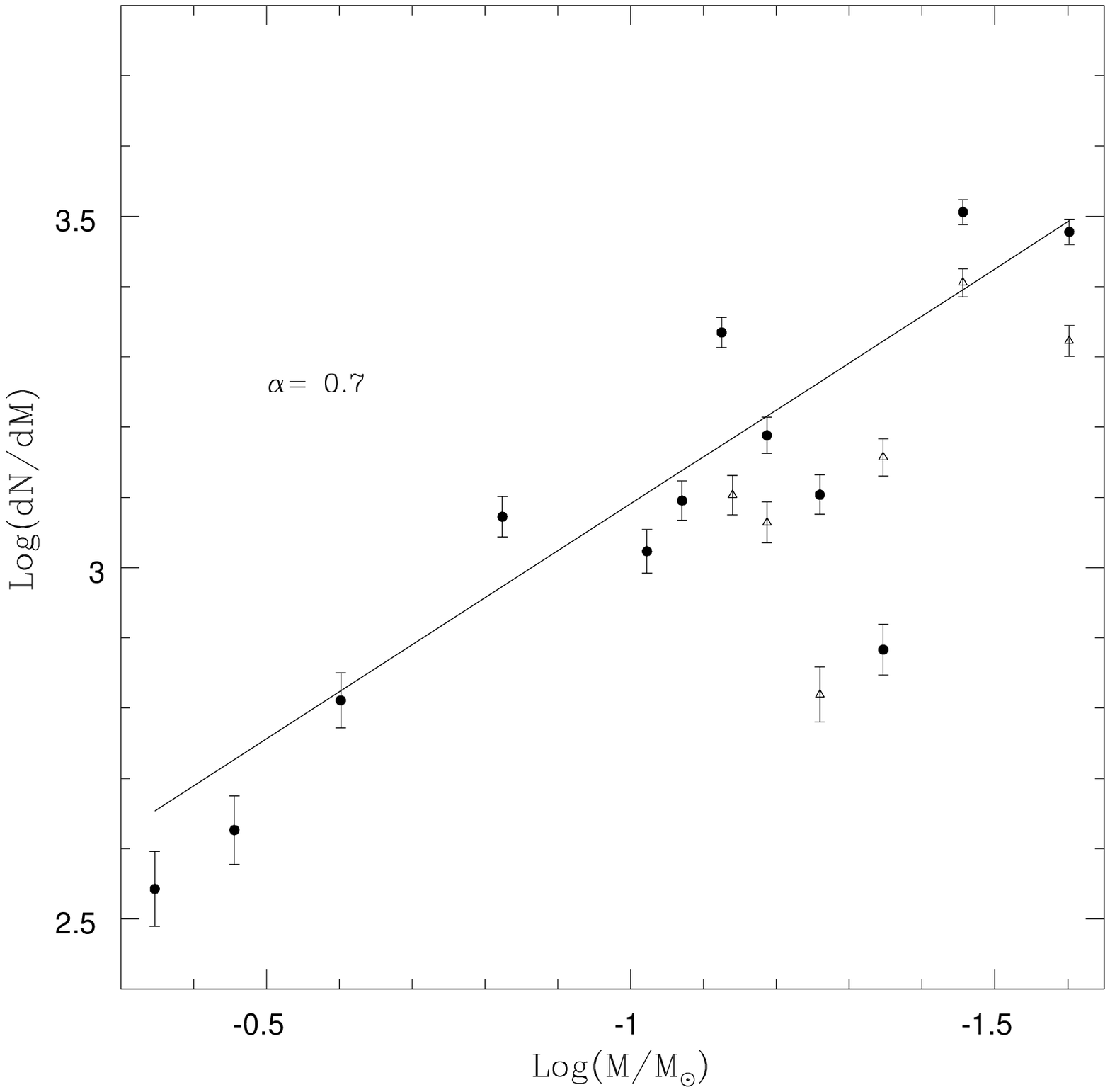}
\figcaption{The mass function derived for IC 348. The y-error
bars are the $\sqrt{N}$ errors of the counting statistics. The filled
circles are data points derived using the models of Baraffe et al. (1998) 
and the open triangles represent data points from the models of
Chabrier et al. (2000). This figure also plots the least squares straight
line fit to the data points.}
\end{figure}}

The mass function derived here is consistent with that derived by
Najita et al. (2000) and Luhman et al. (1998). Using the tracks from
D'Antona \& Mazzitelli (1997), Luhman et al. (1998) derive a mass
function which, in logarithmic units, slowly rises from the HBML to
$\sim$0.25 $M_{\odot}$ with a slope of $\sim -0.4$. Converting their
results to our system (Eqn. 1 \& 2) yields a value of $\alpha$ = 0.6.
Najita et al. (2000) derive the mass function using the models of
Baraffe et al. (1998) as well the evolutionary tracks of D'Antona \&
Mazzitelli (1997). They derive mass functions with $\alpha$ = 0.5 and
$\alpha$ = 0.4 respectively for the two models. 

{\begin{figure}
\plotone{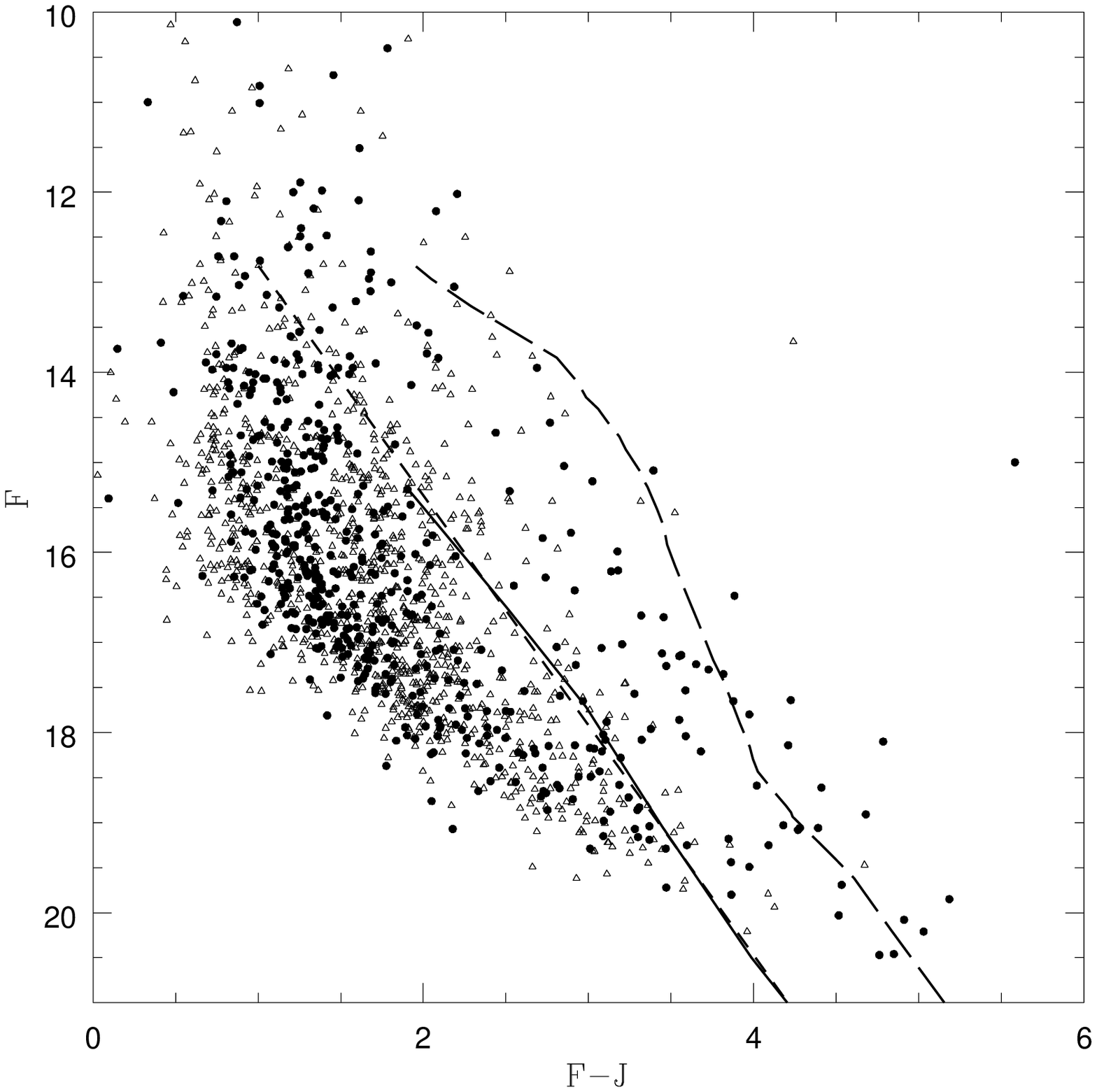}
\figcaption{Same as Fig. 5 but for $\sigma$ Orionis
showing every fifth point of the cluster and the control sample. Here,
the plotted isochrone is for 3 Myr from Baraffe et al. (1998).}
\end{figure}}

\subsection{$\sigma$ Orionis}

We have assumed a mean age of 3 Myr from the age estimates of B\'{e}jar
et al. (1999) for $\sigma$ Orionis. Lee (1968) finds a low extinction
of $E(B-V)$ = 0.05 for the multiple star $\sigma$ Orionis which
suggests that the reddening of the associated cluster is small. We also
assume this value of $E(B-V)$ in our analysis as the value of the
foreground extinction for the cluster. We have covered an area of
$\sim$ 0.8 $deg^{2}$ centered around the star $\sigma$ Orionis. Taking
the distance to be 352 pc (B\'{e}jar et al. 1999), we derive the
following criteria for field star rejection and selection of low mass
members

\begin{equation} F < 2.55(F-J) + 10.26 \end{equation}
\begin{equation} F - K \geq 4.35 \end{equation}
\begin{equation} K \geq 11.17 \end{equation}
\begin{equation} J - K \geq 0.94 \end{equation}
\begin{equation} H - K \geq 0.20. \end{equation}

{\begin{figure}
\plotone{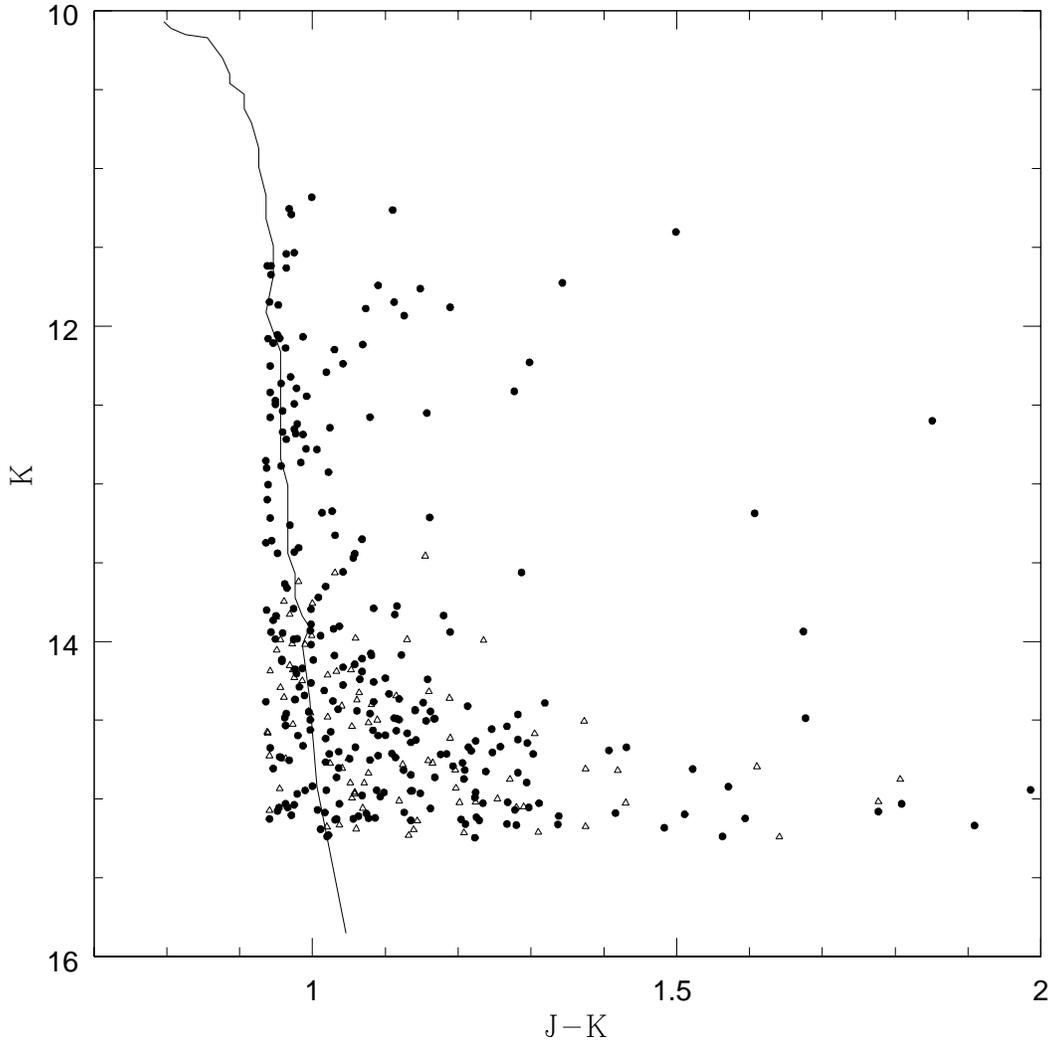}
\figcaption{Same as Fig. 6 but for $\sigma$ Orionis.}
\end{figure}}

Fig. 9-11 show the different CMDs and color-color plots
for $\sigma$ Orionis. As is evident from the figures, the
contamination for this cluster is higher than that
of IC 348. The possible reason might be the location in the Orion
Complex and inadequate correction for variable extinction in the
cluster and the control field.

{\begin{figure}
\plotone{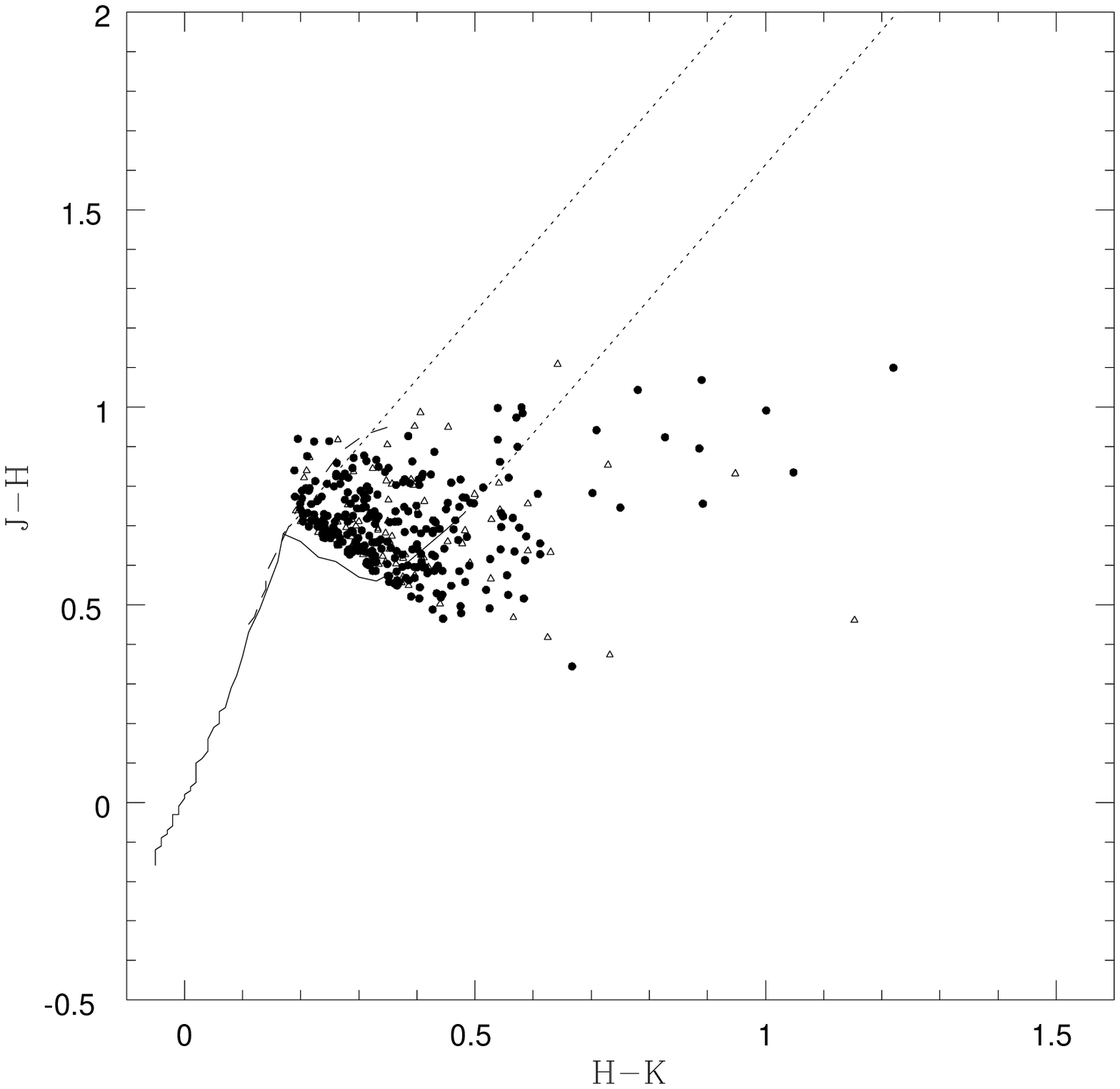}
\figcaption{Same as Fig. 7 but for $\sigma$ Orionis.}
\end{figure}}
 
The derived mass function is shown in Fig. 12. As seen in the figure
there is a sudden rise in the mass function at 0.045 $M_{\odot}$, 
which could be because of the large contamination from the background
sample. The lowest mass bin is affected by the limiting magnitude of
the survey. Hence, these two lowest mass bins were excluded in
deriving the slope of the mass function. The resultant mass function
has a slope of $\alpha$ = 1.2$\pm$0.2. B\'{e}jar et al. (2001) have
previously derived the mass function for this cluster in the mass range
from 0.2 $M_{\odot}$ to 0.013 $M_{\odot}$ based on deep photometry in
I,Z,J and K bands. They obtain a slope of $\alpha$ =0.8$\pm$0.4
assuming the age to be 5 Myr. Our study, which covers a larger area and
includes photometry in all the 2MASS near-IR bands, is consistent with
the results of B\'{e}jar et al. (2001). 

{\begin{figure}
\plotone{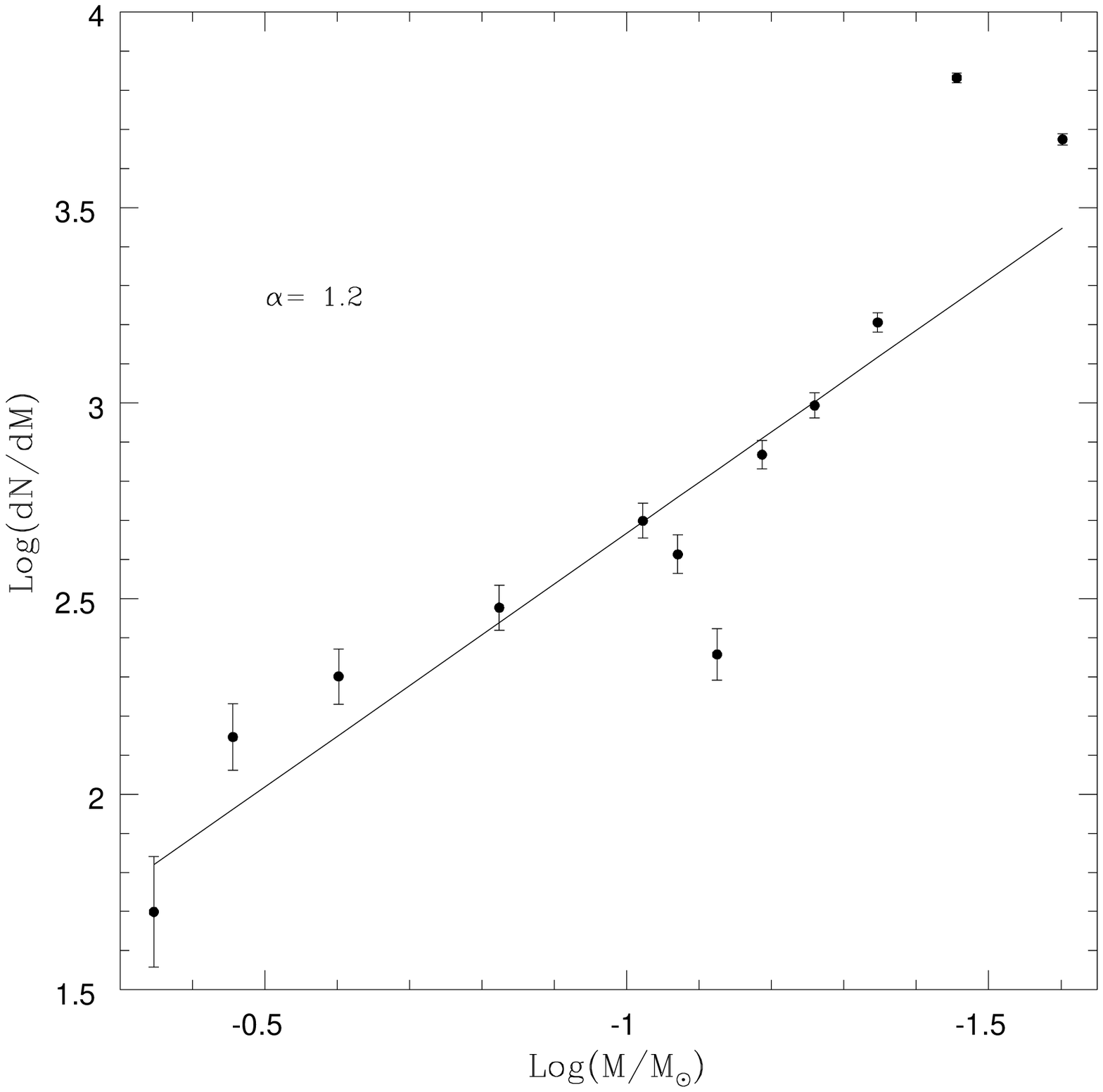}
\figcaption{The mass function derived for $\sigma$ Orionis.}
\end{figure}}

\subsection{Pleiades}

For Pleiades, we have selected four fields of 1 deg radius each and
centers offset by a degree from the cluster center towards the north,
south, east and west. These four fields are then combined together to
one field with an effective radius of 1.5 deg centered around the
cluster center and hence amounts to a total area of $\sim$7 $deg^{2}$.
In contrast to many of the previous studies, our study thus covers a
large area and includes the photometry in optical as well as the
near-IR wavelengths. The control fields are generated in similar
fashion. As compared to IC 348 and $\sigma$ Orionis, Pleiades is an
older cluster with an age estimate varying between $100 - 125$
Myr. In our analysis, we assume an age of 100 Myr (Bouvier et al.
(1998), a distance of 125 pc (Bouvier et al. 1998) and a redenning of
$A_{v}$=0.12 (Crawford \& Perry 1976) for the cluster. Fig. 13
shows the $F$ versus the $(F-J)$ diagram for Pleiades. Because of the
high galactic latitude of this cluster one sees a better separation
between the field and the cluster members unlike that in the other two
clusters. As explained in section 4.2, we have used the locus of the
Baraffe model to set a criterion for rejection of the field population.
A 4th order polynomial best fits the theoretical isochrone for a 100
Myr cluster. The other criteria are derived analogous to the previous
two clusters. The resultant criteria are as follows.

{\begin{figure}
\plotone{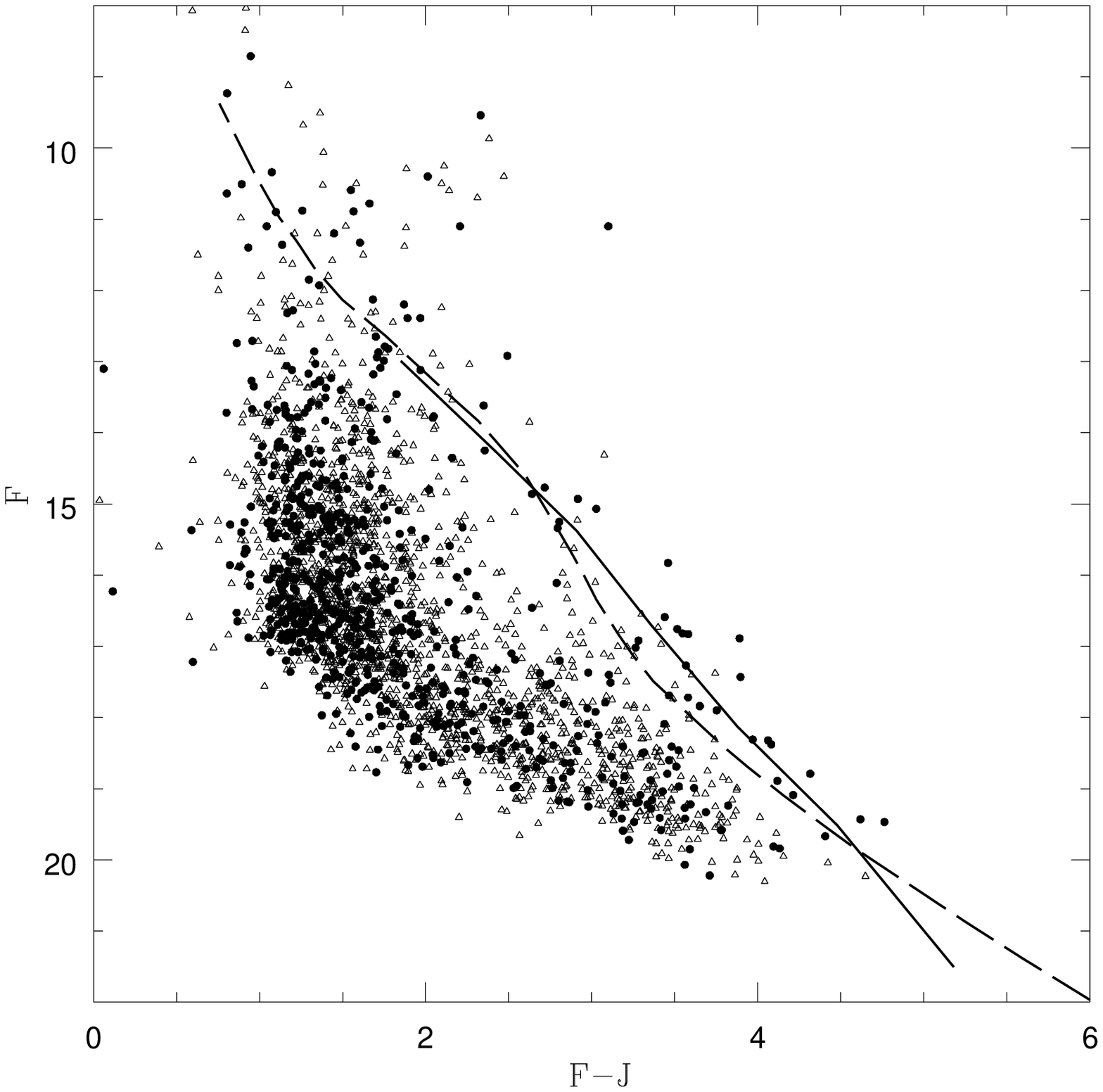}
\figcaption{Same as Fig. 5 but for Pleiades showing
every fifteenth point of the cluster and the control fields. The
model isochrone here is for 100 Myr and the fit to this model
isochrone is given by Eqn. 13.}
\end{figure}}

{\begin{figure}
\plotone{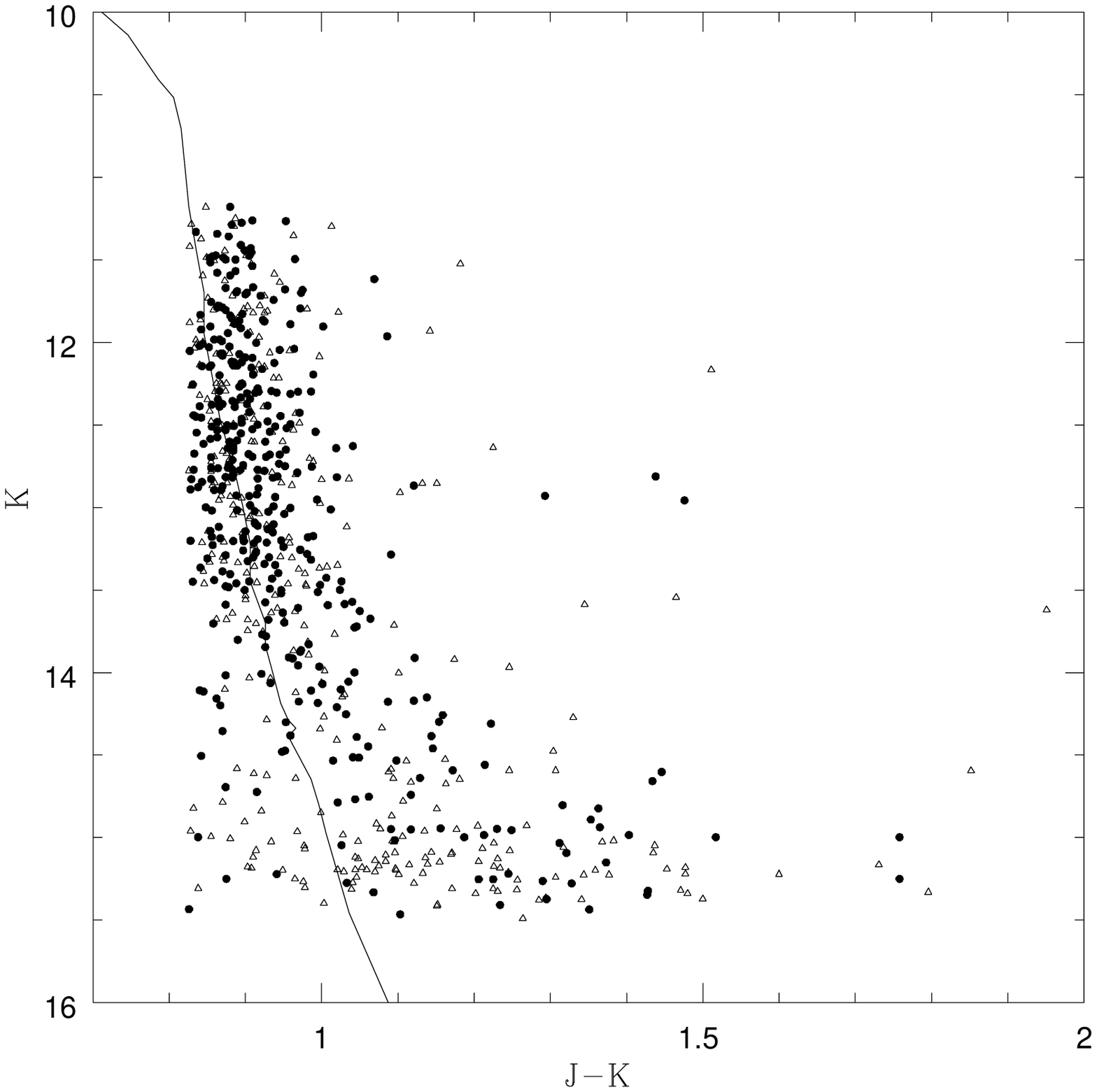}
\figcaption{Same as Fig. 6 but for Pleiades.}
\end{figure}}

\begin{equation} F < 7.93+1.99(F-J)+0.56(F-J)^{2}-0.13(F-J)^{3}+0.01(F-J)^{4}
\end{equation}
\begin{equation} F - K \geq 3.42 \end{equation}
\begin{equation} K \geq 11.17 \end{equation}
\begin{equation} J - K \geq 0.83 \end{equation}
\begin{equation} H - K \geq 0.22 \end{equation}

Fig. 13-15 show the color -- magnitude and the color
-- color plots for Pleiades.

{\begin{figure}
\plotone{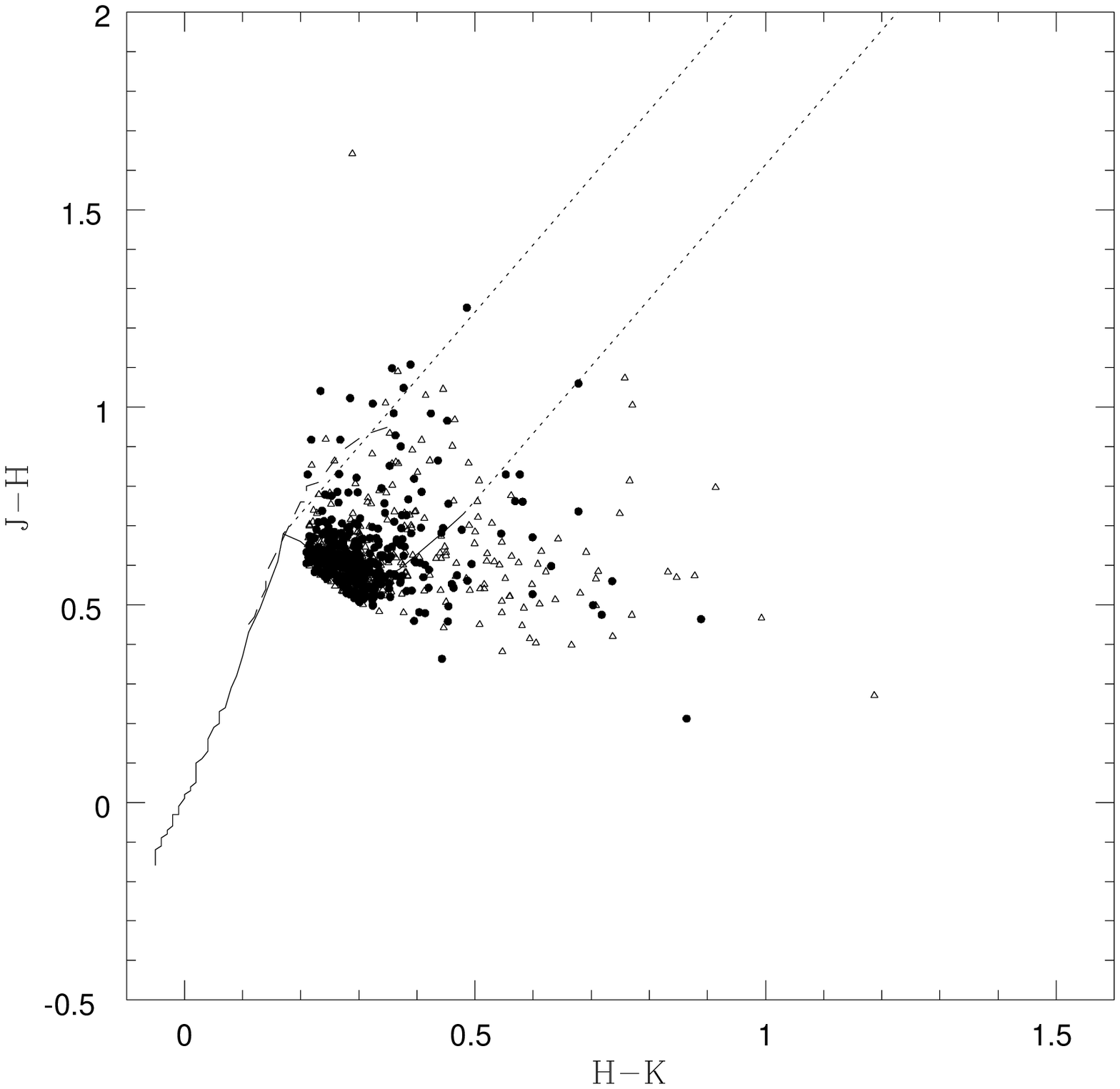}
\figcaption{Same as Fig. 7 but for Pleiades.}
\end{figure}}

Fig. 16 shows the derived mass function. The linear fit to the data
points yields a value of $\alpha$=0.5. This is in good agreement with
the results of Bouvier et al. (1998) who derive a mass function which
still rises below the HBML with $\alpha$=0.6. Mart\'{i}n et al. (1998),
from a compilation of independent photometric survey covering
different areas of the cluster, derive a steeper slope in the mass
range 0.04 -- 0.25 $M_{\odot}$ with $\alpha$=1.0$\pm$0.15. Hambly et
al. (1999) derive a value of $\alpha$=0.7 from a $R$ and $I$ survey
covering an area of 6 $\times$ 6 degrees centered on Pleiades. It is
worth noting here that Pleiades is a relatively older cluster, where 
the mass segregation and the escape mechanism of low-mass stars
described earlier might have played a role. This may be the reason for
the flatter slope compared to the other two clusters, and may also also
explain the observed discrepancies in various estimates of the slopes
indicating that the mass function may not be uniform over the entire
region of the cluster.

{\begin{figure}
\plotone{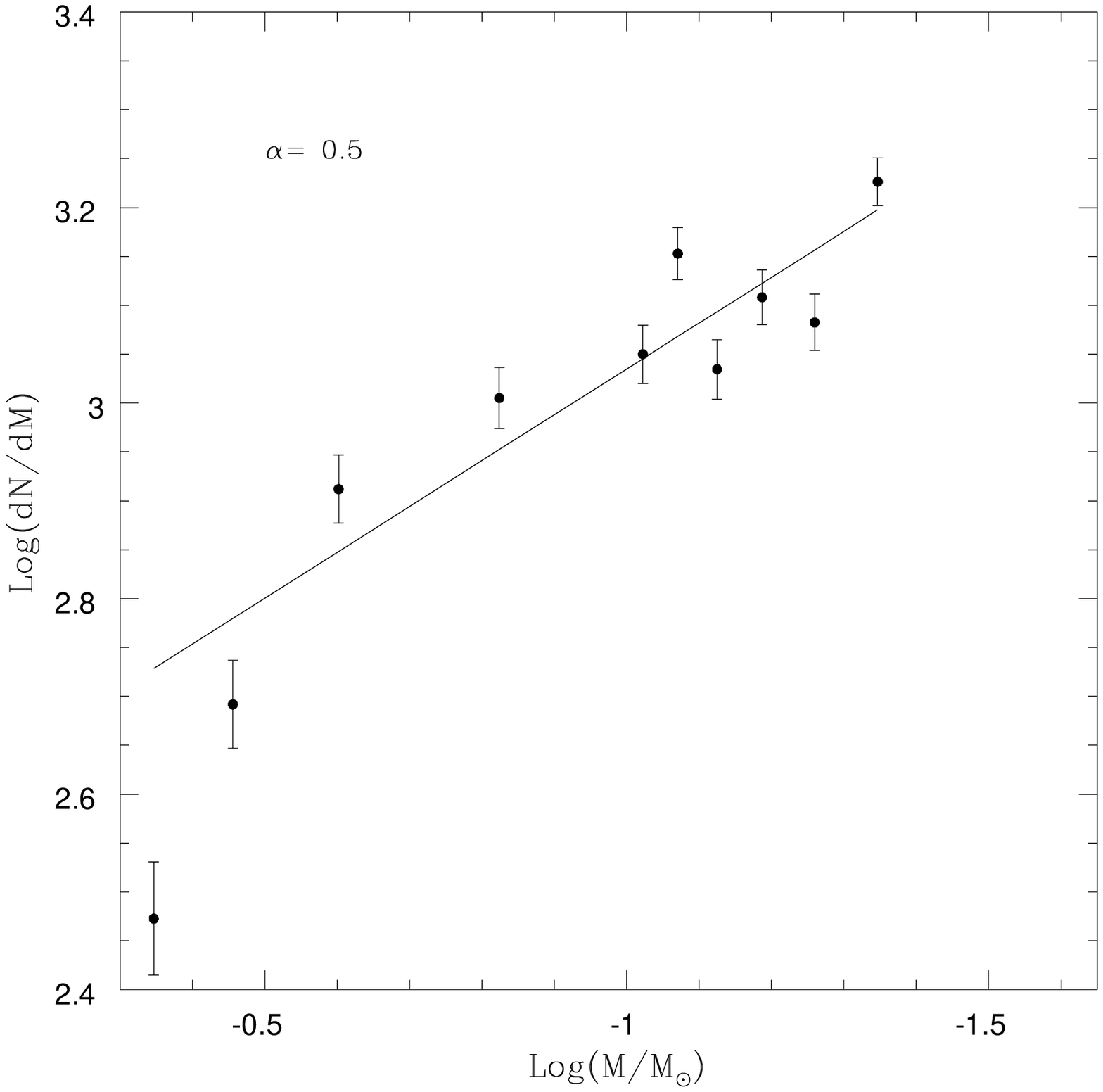}
\figcaption{The mass function derived for Pleiades.}
\end{figure}}

\section{Discussion and Conclusion}

Recent surveys have found a significant population of low mass stars,
brown dwarfs and planetary mass objects in young open clusters. We have
adopted a statistical approach to determine the mass spectrum, $dN/dM
\propto M^{-\alpha}$, of objects in the mass range 0.5$M_{\odot}$ to 
0.025-0.055$M_{\odot}$, using the data from the recently released 
2MASS and the GSC catalogues. Unlike some of the previous studies, our
study makes use of both the optical data as well as the near-IR data.
Since these datasets cover a large portion of the sky, they 
allow us to study the
entire area covered by each cluster. As a result, the areas covered in
our study are generally larger than the areas covered in most of the
previous studies. These datasets also allow us to apply a statistical
approach to efficiently subtract the background contribution using
several control fields close to the cluster. 

We carried out a detailed study for IC 348. Fo this cluster, we derived
the mass functions using the solar metalicity model of Baraffe et al.
(1998) and the dusty models of Chabrier et al. (2000), both of which
gave very similar results. We also compared the confirmed low-mass
members from Luhman (1999) with our isolated low-mass candidates which
further strengthened the validity of this technique. We then used the
same technique to $\sigma$ Orionis and Pleiades. The resultant slopes
of the mass functions for IC 348, $\sigma$ Orionis and Pleiades are
0.7, 1.2 and 0.5 respectively, with an estimated error of $\pm$0.2. For
IC 348 we have used the mass range from 0.5$M_{\odot}$ to
0.025$M_{\odot}$ in deriving the slope of the mass function whereas for
$sigma$ Orionis and Pleiades, the lowest mass bins correspond to 0.045
$M_{\odot}$ and 0.055$M_{\odot}$, respectively. Taking into
consideration the effect of mass segregation and preferential loss of
low mass members from the cluster, the mass function derived here for
the inner 1.5 deg radius of Pleiades cluster could be considered as a
lower limit to the true mass function. As discussed under individual
sections, the mass functions derived here are in good agreement with
the value derived by other groups based on studies of confirmed low
mass members of these clusters, which demonstrates the consistency of
different approaches. Within the uncertainties, the values of $\alpha$
derived for the IC 348 and $\sigma$ Orionis is in agreement with that
derived for the local sample of low-mass stars (Reid et al. 1999). The
derived slopes imply that the mass spectrum continues to rise well
below the HBML, but the mass functions are appreciably flatter in the
low-mass regime than the Salpeter mass function. The results are
summarized in Table 3.

Taking the Salpeter exponent of 2.35 in the mass range 1 -- 10
$M_{\odot}$, the Chabrier exponent of 1.55 in the mass range 0.5 -- 1
$M_{\odot}$, and the values obtained by us below 0.5, we calculate the
mass contribution to be about 40\% for objects below 0.5 M$_\odot$,
and about 4\% for objects below the HBML of 0.08 M$_\odot$. (Note that
the contributions are not sensitive to the choice of the slope in the
higher mass regime. For example, if we use the value of $\alpha$ as
2.7 for M$> 1 M_\odot$ as derived by Chabrier (2001) instead of the
Salpeter value of 2.35, the mass contributions change only by
$\sim1\%$). Our results are consistent with that of the previous
studies (e.g. B\'{e}jar et al. 2001), and suggest that, although the
low mass stars are at least as numerous as their high mass counter
parts (as seen from figures 5,7 and 9), their contribution to the total
mass is small. The contributions of low-mass objects to the total mass
in the clusters seem to be marginally smaller than that of the low-mass
objects in the local sample (e.g. Reid et al. 1999), but the slope of
the mass function is less steep for the relatively older Pleiades
cluster. This is not surprising since the high-mass stars are likely
to be preferentially lost in an older and mixed population such as the
local sample. 

Follow up spectroscopic observations to confirm the isolated low mass
members of the clusters would further strengthen the results derived
from this purely statistical approach. 

\section*{acknowledgements} \noindent We would like to thank Brian
McLean and Mario Lattanzi (the GSC-II project scientists) for access to
the development version of GSC-II in advance of publication and their
technical comments. We would like to thank the anonymous referee for
valuable comments and suggestions. We are grateful to I. Baraffe and F.
Allard for making the electronic versions of the latest models
available and generating the model isochrones for the non-standard F
passbands. We would also like to thank Neill Reid for useful
discussions. The visit of A. Tej to STScI was supported by a DDRF grant
of STScI awarded to K. Sahu. The Digitized Sky Surveys and the Guide
Star Catalogues were produced at the Space Telescope Science Institute
under US Government grant NAG W-2166; the images of these surveys are
based on photographic data obtained using the Oschin Schmidt Telescope
on Palomar mountain and the UK Schmidt Telescope at Siding Spring. The
2MASS project is a collaboration between The University of
Massachusetts and the Infrared Processing and Analysis Center (JPL/
Caltech), the funding for which is provided primarily by NASA and the
NSF. Research work at PRL is funded by Dept. of Space, Govt of India.

\newpage

\begin{table}
\caption{Mass limits for the three clusters}
\begin{tabular}{ccccc}
\hline
\hline
Passband & Limiting Mag.      & \multicolumn{3}{c}{Corresponding Mass($M_{\odot}$)} \\
         &                    & IC 348 & $\sigma$ Orionis & Pleiades \\
\hline
$J$      & 16.5               & 0.025  & 0.025         & 0.04     \\
$H$      & 15.5               & 0.025  & 0.025         & 0.04     \\
$K$      & 15.0               & 0.025  & 0.025         & 0.04     \\
$F$      & 21.0               & 0.025  & 0.025         & 0.04     \\
\hline
\end{tabular}
\end{table}

\begin{table}{}
\caption{Positions of the Field Centers}
\begin{tabular}{cccc}
\hline
\hline
Fields & R.A. (J2000.0) & Decl. (J2000.0) & Radius \\
       & (hh mm ss)       & (dd mm ss)       & (arcmin)\\
\hline
IC 348    & 03 44 30 & +32 17 00 & 20 \\
Control 1 & 03 49 08 & +31 19 08 & 20\\
Control 2 & 03 44 10 & +33 19 26 & 20 \\
\hline
$\sigma$ Orionis & 05 38 45 & -02 36 00 & 30\\
Control 1        & 05 58 29 & -04 29 48 & 30\\
Control 2        & 05 11 00 & -00 20 00 & 30\\
\hline
Pleiades & 03 47 00 & +24 07 00 & 90\\
Control 1 & 03 18 00 & +26 41 00 & 90\\
Control 2 & 03 05 00 & +24 42 00 & 90\\
\hline
\end{tabular}
\end{table}

\begin{table}{}
\caption{Summary of Results}
\begin{tabular}{ccccc}
\hline
\hline
Cluster	& Age  & Distance  &  Mass Range    & $\alpha$ \\
	& (Myr)& (pc)      & ($M_{\odot}$) &          \\
\hline
IC 348  & 5    & 316       & 0.5 -- 0.035  & 0.7      \\
$\sigma$ Orionis  & 3 & 352   & 0.5 -- 0.045  & 1.1      \\
Pleiades       & 100 & 125    & 0.5 -- 0.055  & 0.5      \\
\hline
\end{tabular}
\end{table}

\end{document}